\documentclass[letterpaper,twocolumn,10pt]{article}
\usepackage{usenix-2020-09}

\PassOptionsToPackage{hyphens}{url}
\usepackage{hyperref}
\usepackage{url}

\usepackage{cite}
\usepackage{amsmath,amssymb,amsfonts}
\usepackage{algorithmic}
\usepackage{graphicx}
\usepackage{subfig}
\usepackage{textcomp}
\usepackage{xcolor}
\usepackage{booktabs}
\usepackage{multirow}
\usepackage{titlesec}
\usepackage{soul}
\usepackage{changepage,threeparttable}

\usepackage{hyperref}
\usepackage[misc]{ifsym}

\titlespacing*{\section}
{0pt}{1.3ex plus 1ex minus .2ex}{1.3ex plus .2ex}

\titlespacing*{\subsection}
{0pt}{1ex plus .2ex}{1ex plus .2ex}

\usepackage{pifont}
\usepackage{colortbl}

\usepackage[T1]{fontenc}
\usepackage{tikz}

\usepackage{appendix}

\usepackage{tabularx}
\usepackage{stfloats}
\usepackage{placeins}
\usepackage{booktabs}
\usepackage{float} % 引入 float 包以支持 [H]
\usepackage{lipsum}
\usepackage{afterpage}
\usepackage{threeparttable}
\usepackage{amsmath}

\usepackage[subtle]{space/savetrees} %CSPACE
\renewenvironment{quote}{%
  \list{}{%
    \leftmargin0.2cm   % this is the adjusting screw
    \rightmargin\leftmargin
    \topsep=0.5pt
  }
  \item\relax
}
{\endlist}

\if0
\newcommand{\xw}[1] {{\color{blue}#1}}
\newcommand{\jingtao}[1] {{\color{blue}#1}}

\newcommand{\TODO}[1] {{\color{blue}TODO: #1}}

\fi

\newcommand{\xwnew}[1] {{\color{black}#1}}

\newcommand{\xw}[1] {{\color{black}#1}}
\newcommand{\jingtao}[1] {{\color{black}#1}}

\newcommand{\TODO}[1] {{\color{black}TODO: #1}}

\newcommand{\para}[1]{\vspace{2pt}\noindent\textbf{#1.~}}
\newcommand{\ignore}[1]{}

\usepackage[T1]{fontenc}
\usepackage{tikz}

\usepackage{enumitem}

\begin{document}

%\title{Four Years After Paternalistic Privacy Regulation Enforcement: A Qualitative Analysis of China's Special Privacy Rectification}

%\title{Four Years After Paternalistic Privacy Regulation Enforcement: A Qualitative Analysis of China's Special Privacy Rectification}

%\title{Uncovering Privacy Law Enforcement under Centralized Governance: A Qualitative Analysis of China's Special Privacy Rectification}

\title{Privacy Law Enforcement Under Centralized Governance:\\ A Qualitative Analysis of Four Years' Special Privacy Rectification Campaigns}

\author{
{\rm Tao Jing$^{1,2,}\footnotemark[1]$~~, Yao Li$^{3}$, Jingzhou Ye$^{3}$, Jie Wang$^{1,2,\textrm{\Letter},}\footnotemark[1]~~$, and Xueqiang Wang$^{3}$}
\and
% \small
% $^1$ Hubei Key Laboratory of Distributed System Security\\
% \small
% $^2$ Hubei Engineering Research Center on Big Data Security\\
% \small
$^1$ School of Cyber Science and Engineering, Huazhong University of Science and Technology\\
% \small
$^2$JinYinHu Laboratory\\
% \small
$^3$ University of Central Florida\\
% \small
$^\textrm{\Letter}$ Corresponding author: wangjie\_s@hust.edu.cn\\
% copy the following lines to add more authors
% \and
% {\rm Name}\\
%Name Institution
} % end author
% make the title area
\maketitle
% \clearpage
\renewcommand{\thefootnote}{\fnsymbol{footnote}}
% \footnotetext[1]{Corresponding author: wangjie\_s@hust.edu.cn. }
\footnotetext[1]{Hubei Key Laboratory of Distributed System Security, Hubei Engineering Research Center on Big Data Security, School of Cyber Science and Engineering, Huazhong University of Science and Technology.}
\begin{abstract}
\xwnew{In recent years, major privacy laws like the GDPR have brought about positive changes. However, challenges remain in enforcing the laws, particularly due to under-resourced regulators facing a large number of potential privacy-violating software applications (apps) and the high costs of investigating them.}
Since 2019, China has launched a series of privacy enforcement campaigns known as Special Privacy Rectification Campaigns (SPRCs) to address widespread privacy violations in its mobile application (app) ecosystem. 
\xwnew{Unlike the enforcement of the GDPR, SPRCs are characterized by large-scale privacy reviews and strict sanctions, under the strong control of central authorities.
In SPRCs, central government authorities issue administrative orders to mobilize various resources for market-wide privacy reviews of mobile apps.
They enforce strict sanctions by requiring privacy-violating apps to rectify issues within a short timeframe or face removal from app stores.}
While there are a few reports on SPRCs, the effectiveness and potential problems of this campaign-style privacy enforcement approach remain unclear to the community. 

\ignore{When General Data Protection Regulation (GDPR) is enforced, Data Protection Authorities (DPAs) investigates user complaints and industry reports of privacy breaches, and issue warnings and fines accordingly. However, due to the reactive nature of these supervisory agencies, the enforcement of GDPR compliance is often delayed, e.g., compliance after investigations.}
\ignore{The enforcement of major privacy laws, such as the General Data Protection Regulation (GDPR), relies on supervisory agencies to investigate potential privacy violations and apply penalties in a largely autonomous manner, guided by the general privacy principles of the laws.
While this approach is good for consistent and continuous enforcement of privacy laws, it may not rapidly improve compliance in ecosystems with widespread privacy violations.
Four years ago, China introduced a distinctive campaign-style enforcement of privacy laws known as Special Privacy Rectification Campaigns (SPRCs) with the goal of quickly and effectively addressing the serious privacy risks posed by Chinese mobile apps.
Nevertheless, the effectiveness and potential problems of this enforcement approach are still unclear to the community.}

\xw{In this study, we conducted 18 semi-structured interviews with app-related engineers involved in SPRCs to better understand the campaign-style privacy enforcement. 
Based on the interviews, we reported our findings on a variety of aspects of SPRCs, such as the processes that app engineers regularly follow to achieve privacy compliance in SPRCs, the challenges they encounter, the solutions they adopt to address these challenges, and the impacts of SPRCs, etc. 
We found that app engineers face a series of challenges in achieving privacy compliance in their apps. For example, they receive inconsistent app privacy review reports from multiple app stores and have difficulties confirming the issues flagged by these reports; they also lack institutional support for studying privacy laws, self-validating privacy compliance of their apps, communicating effectively between multiple stakeholders, and ensuring fairness in accountability when privacy non-compliance occurs.
Furthermore, we found that while SPRCs have introduced several positive changes, there remain unaddressed concerns, such as the potential existence of circumvention techniques used to evade app privacy reviews.}
\end{abstract}
\section{Introduction}
\label{sec:intro}

\begin{quote}
\noindent\textit{``Laws without enforcement are just good advice.''}

\hspace*{\fill}- Abraham Lincoln
\vspace{4pt}
\end{quote}

\xwnew{Recent studies have shown that the enforcement of privacy laws has led to a variety of positive changes, such as improved privacy policies~\cite{linden2020privacy,zaeem2020effect}, reduced use of tracking cookies~\cite{libert2018changes}, and even increased company revenues~\cite{cao2024regulation}.
However, challenges still exist in enforcing the laws, particularly due to under-resourced regulators \cite{egelman2023informing} and the high costs of investigating privacy-violating software applications (apps).
For example, the Federal Trade Commission (FTC), which enforces the Children’s Online Privacy Protection Act (COPPA) \cite{coppa}, takes an average of 294 days to complete an investigation \cite{ftcdays}, a lengthy process that includes evidence collection, violation assessment, and court proceedings, etc. 
Considering the large number of apps available (in the millions on mobile platforms~\cite{statista2024}), many apps may not undergo external privacy reviews, even though they pose privacy risks~\cite{kollnig2021fait,reyes2018won}.}

\ignore{As organizations prioritize revenue tied to personal data collection and lack the professional resources for privacy compliance\cite{amos2021privacy,xiao2022lalaine,reyes2018won,samarin2023lessons}, it is impractical to expect them to self-regulate to satisfy the law enforcement.
Consequently, a proactive mechanism seems to be what it needs to enforce organizations to protect user privacy.}
%the exploration of alternative approaches to achieving widespread privacy compliance, and understanding the new opportunities and challenges they bring, has become unprecedentedly important. 

%\para{Special privacy rectification campaigns (SPRCs)}
%
\ignore{To tackle the widespread privacy violations in its mobile app ecosystem, since October 2019, China has launched a series of privacy enforcement campaigns known as the Special Privacy Rectification Campaigns (SPRCs)~\cite{sprc2019,sprc2020,sprc2021,sprc2023}.
}
\xw{Since October 2019, China has launched a series of privacy enforcement campaigns known as the Special Privacy Rectification Campaigns (SPRCs)~\cite{sprc2019,sprc2020,sprc2021,sprc2023} to tackle the widespread privacy violations in its mobile app ecosystem.
\xwnew{Unlike the lengthy investigation process, SPRCs are characterized by large-scale and strict sanctions for privacy governance, under the strong control of central authorities such as the Ministry of Industry and Information Technology (MIIT)~\cite{miit}.}
These authorities issue administrative orders that specify detailed privacy compliance requirements, and mobilize necessary resources (including major app stores and third-party privacy certifiers) to conduct comprehensive, market-wide privacy reviews of mobile apps.
Apps found to have privacy violations must be rectified by their providers within a defined timeframe (e.g., five business days), or the apps risk being listed on public privacy bulletins by the MIIT or removed from app stores~\cite{sprc2019,sprc2020,sprc2021,sprc2023}.
According to available data as of June 2022, MIIT has conducted intensive privacy reviews for over 3.22 million apps, resulting in the removal of at least 3,000 apps from app stores~\cite{rectifyingresults}.}

However, although there are a few government reports on SPRCs~\cite{rectifyingresults,tc260js2pdf}, \xwnew{the effectiveness and potential problems of such a campaign-style enforcement approach, characterized by large-scale privacy reviews and strict sanctions}, remain largely unknown to both industry and academia.
Understanding this approach is crucial not only for evaluating the outcomes of China's investment in privacy law compliance but also for guiding future steps in this area.
Moreover, in the long term, analyzing this campaign-style enforcement approach can provide valuable insights into alternative methods of privacy law enforcement and potentially improve privacy compliance efforts in other countries.

%\para{A interview study for analyzing SPRCs}
%
This study qualitatively analyzes SPRCs to better understand the effectiveness and potential problems in these large-scale enforcement efforts. 
Specifically, we plan to answer the following research questions: 

\begin{itemize}[left=0pt, label={}, itemsep=3pt]
    \item \textit{RQ1: What is the workflow that privacy stakeholders regularly follow in SPRC?}
    \item \textit{RQ2: What challenges did app developers encounter in achieving app privacy compliance in SPRCs?}
    \item \textit{RQ3: What solutions have the app developers adopted to address these challenges?}
    \item \textit{RQ4: What are the overall impacts of the SPRCs?}
\end{itemize}

To answer these questions, we conducted a semi-structured interview study involving 18 app-related engineers who have been involved in SPRCs.
These participants consist of app developers, technical leads, security engineers, and test engineers, all of whom have experience in ensuring privacy compliance with SPRCs for mobile apps.
Based on these interviews, we report the following key insights: 

\begin{itemize}
    \item \textit{Inconsistencies manifest in a variety of aspects of SPRCs.} For instance, an app may receive different privacy review reports from multiple app stores, and these reports can also be different to the developers self-test results, which cause frustration for app developers. Also, app stores have differing definitions of sensitive data with app developers, and they treat popular apps more strictly than unpopular apps during app review. 
    \item \textit{There is a lack of institutional support from the app providers for privacy compliance.} App developers reported that, to achieve privacy compliance, they need more support for studying privacy laws, more resources to self-validate privacy compliance, more support for communication, and fairness in accountability. 
    \item \textit{SPRCs result in both positive changes and concerns regarding privacy compliance.} Overall, participants reported that SPRCs reduced the number of privacy-invading apps and increased awareness of the significance of privacy among app engineers. However, they expressed particular concern about the existence of circumvention techniques used to evade app privacy reviews.
\end{itemize}

\section{Background}
\label{sec:background}

\subsection{Enforcement of EU and US Privacy Laws} 
The General Data Protection Regulation (GDPR)~\cite{gdprLink}, which came into effect on May 25, 2018, is a comprehensive legal framework designed to protect the privacy rights of individuals in the European Union (EU). 
\xwnew{To enforce the GDPR, Data Protection Authorities (DPAs) in each EU member state undertake various tasks to monitor organizational compliance, investigate potential violations, promoting public awareness, etc. 
The investigation process, in particular, can be time-consuming~\cite{noyb_article77} due to the need for comprehensive data reviews, legal assessments, and the detailed procedural steps required for potential court proceedings.}
Privacy legislation in the US is based on both federal and state privacy laws. The enforcement of each privacy law is similar to the GDPR in that it relies on enforcement actions by federal or state supervisory agencies, under the guidance of privacy laws.
Examples are the Children’s Online Privacy Protection Act (COPPA)\cite{coppa}, enforced by the Federal Trade Commission (FTC)\cite{ftc}, and the California Consumer Privacy Act (CCPA)\cite{ccpa}, enforced by the California Attorney General's office\cite{cagoffice}.

\subsection{Enforcement of Chinese Privacy Laws}
\label{subsec:background_chinese}

\xwnew{Unlike the enforcement of GDPR and US privacy laws, China adopts a different approach by enforcing its privacy laws through nationwide campaigns, similar to how it enforces other laws~\cite{wang2021legal,wang2024campaign,van2016campaign,yin2023centralized,wang2020politics,zhang2024imbalance,zheng2021application}.
In recent years, Chinese government agencies have launched several privacy enforcement campaigns to conduct large-scale privacy reviews of mobile apps, imposing strict sanctions based on various privacy laws, provisions, and guidelines.}

\begin{figure}[!t]
\centering
\includegraphics[width=5cm]{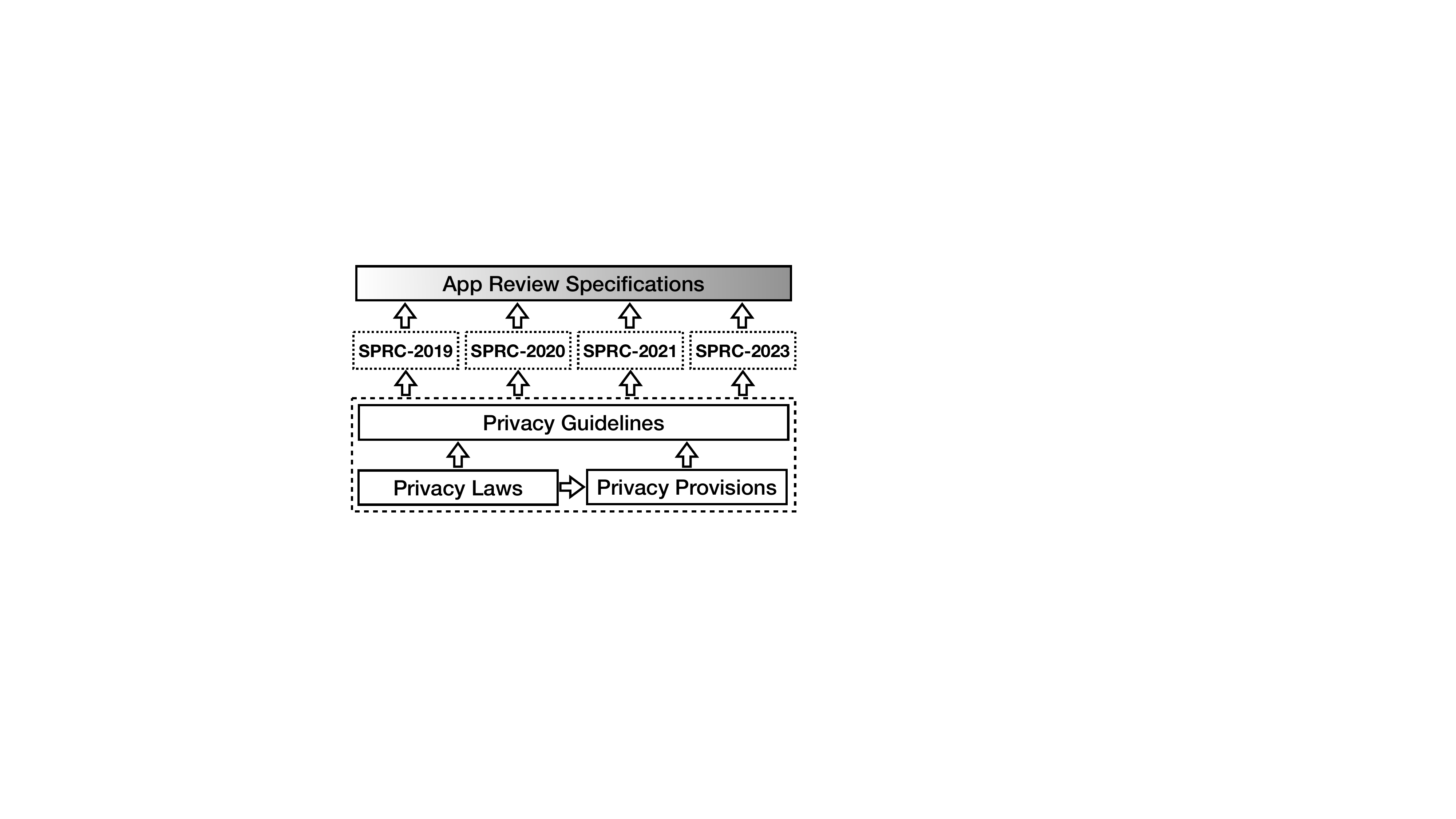}
\caption{Hierarchy of privacy laws, provisions, guidelines, and app review specifications}
\label{fig:law-hierarchy} 
\end{figure}

\para{Chinese privacy laws, provisions, and guidelines}
There are several Chinese privacy laws that outline general privacy principles in a manner similar to GDPR, such as the Cybersecurity Law (2016)\cite{cybersecuritylaw}, the Regulations on Telecommunications (2000)~\cite{regulationsontelecommunications}, and the Personal Information Protection Law (PIPL, 2021)\cite{pipl}, etc.
Based on these laws, Chinese government agencies, such as the Ministry of Industry and Information Technology (MIIT), issue provisions to address specific details of the general privacy principles.
For example, the ``Provisions on Protecting the Personal Information of Telecommunications and Internet Users'' (MIIT Order No. 24, 2013)~\cite{miit2013no24} and the ``Provisions on the Scope of Necessary Personal Information for Common Types of Mobile Applications'' (CAC Order No. 14, 2021)~\cite{cac2021no14} specify requirements for protecting the personal information and rights of Internet users, including mobile app users.
Besides the privacy laws and provisions, government agencies can also release privacy guidelines that offer more practical guidance on interpreting or implementing laws and provisions.
For example, on December 30, 2019, MIIT issued a privacy guideline in MIIT Secret [2019] No. 191, titled ``Means for Determination of Violations of Laws and Regulations in Apps' Collection and Use of Personal Information'' (\textit{Guideline-2019})~\cite{guideline2019}.
Although the guideline is not legally binding, it has evolved into a de facto standard used by stakeholders to assess privacy violations in apps.
Figure~\ref{fig:law-hierarchy} shows the relationship between privacy laws, provisions, and guidelines. 

\para{Special Privacy Rectification Campaigns (SPRCs)}
%
% 电信和互联网用户个人信息保护规定, 常见类型移动互联网应用程序必要个人信息范围规定
%
\xw{While agencies like MIIT also support community-based efforts such as responding to individual complaints and public reports, their predominant enforcement efforts so far have focused on launching nationwide privacy campaigns, often referred to as \textit{Special Privacy Rectification Campaigns} (SPRCs)}.
Specific to the enforcement on mobile platforms, MIIT, in collaboration with three other government agencies including the Cyberspace Administration of China (CAC), has initiated four SPRCs that have spanned over four years.
In October 2019, MIIT launched an SPRC (\textit{SPRC-2019}~\cite{sprc2019}) by issuing an administrative notice to stakeholders aimed at rectifying mobile apps' infringement of users' privacy rights and interests.
Later, in August 2020, July 2021, and February 2023, MIIT revitalized the program with another three SPRCs, \textit{SPRC-2020}~\cite{sprc2020}, \textit{SPRC-2021}~\cite{sprc2021}, and \textit{SPRC-2023}~\cite{sprc2023}, respectively, to further enhance the objectives set forth in prior SPRCs.
% https://www.gov.cn/zhengce/zhengceku/2019-11/13/content_5451505.htm， 工业和信息化部关于开展APP侵害用户权益专项整治工作的通知，2019
% https://www.gov.cn/zhengce/zhengceku/2020-08/02/content_5531975.htm，工业和信息化部关于开展纵深推进APP侵害用户权益专项整治行动的通知，2020
% 《工业和信息化部关于开展互联网行业市场秩序专项整治行动的通知》（工信部信管函[2021]165号）
% https://www.gov.cn/zhengce/zhengceku/2023-03/02/content_5744106.htm，工业和信息化部关于进一步提升移动互联网应用服务能力的通知，2023
%
Note that \ignore{\textit{SPRC-2019} was launched even before the first privacy guideline on mobile apps, \textit{Guideline-2019}, was released, so much of the \textit{SPRC-2019} is based only on high-level laws and provisions. 
Also, }the newer SPRCs do not disable but rather complement prior SPRCs by addressing a broader scope of problems or by prioritizing the types of privacy violations that warrant more attention.
For example, SPRC-2019 requires that app developers and distribution platforms (i.e., app stores) detect and rectify eight types of privacy violations related to the collection, use, and sharing of personal information, as well as the ease of deleting user accounts.
SPRC-2020 also covers these violations and expands the requirements to include emerging violations, such as in personalized ads and deceptive privacy practices, and involves new stakeholders, such as third-party SDK developers.

\begin{figure}[!t]
\centering
\includegraphics[width=8.5cm]{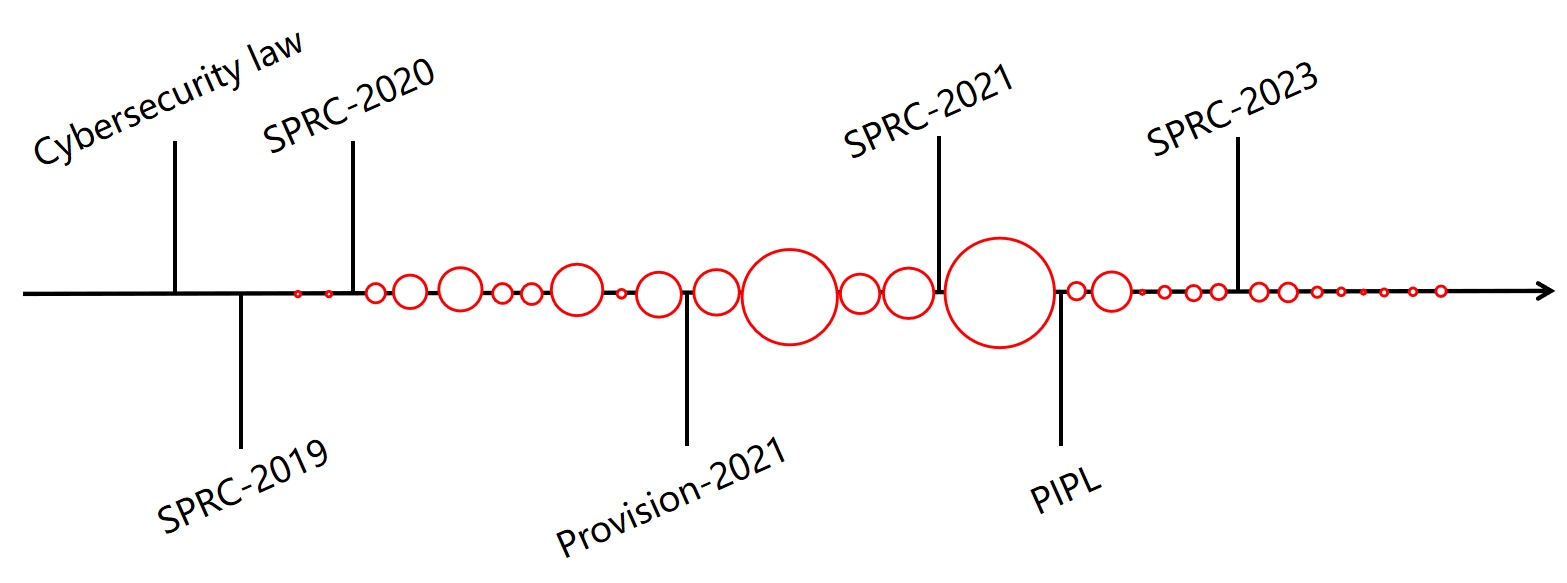}
\caption{Timeline of SPRCs and privacy bulletins issued by government agencies. Each circle (\textcolor{red}{\scriptsize\textcircled{}}) represents a bulletin, with the size of the circle indicating the number of reported apps with privacy violations.}
\label{fig:timeline} 
\end{figure}

According to administrative notices~\cite{sprc2019,sprc2020,sprc2021,sprc2023}, the SPRCs conduct two types of app privacy reviews: government agencies perform privacy reviews on existing apps in app stores, either directly or via third-party privacy certification services (certifiers), and major app stores are required to conduct privacy reviews for newly submitted apps.
Typically, privacy certifiers and app stores develop their own sets of app review specifications based on the administrative notices of SPRCs. 
If an app is found to have privacy violations, the app provider must rectify them; otherwise, there is a risk of the app being delisted or facing fines.
Notably, the SPRCs are not short-term campaigns but represent long-term, ongoing efforts that have extended for more than four years until now.
To date, the agencies have issued warnings of delisting through 36 public privacy bulletins on their websites (such as~\cite{cacfirstbatch2019}), akin to walls of shame, with each bulletin listing hundreds (or at least dozens) of privacy-violating apps. Figure~\ref{fig:timeline} shows the SPRCs and release time of the privacy bulletins. 

\section{Methodology}

\label{sec:methodology}

In fall 2023, we conducted a semi-structured interview study with 18 participants who have been involved in the privacy compliance with SPRCs, to explore their experience and perception of China's large-scale SPRCs, as well as their challenges and solutions.  
\ignore{\TODO{summarize the RQs here}. In this section, we describe the study procedures.}

\subsection{Recruitment of Participants}

\xw{We adopted convenience sampling to recruit participants who have been involved in SPRCs on WeChat~\cite{wechat} -- one of the largest social media in China.}
\jingtao{
Through our personal networks, we joined five WeChat groups with names such as ``App Privacy Compliance Discussion Group''\ignore{of interest}.
}
These groups are composed of a total of 1,600 members who have worked on privacy compliance from different industries and companies. The group members discuss a variety of privacy compliance issues, such as how to use the detection tools, how to address noncompliance, and what is the latest reported noncompliance and the updated privacy review specifications.
\jingtao{To reach a more diverse sample, we were then referred to another four groups by our participants. However, after reviewing the chat history for one week, there was no privacy compliance-related content found in the these four groups, and thus we did not recruit from these groups.}
We posted recruitment information in the WeChat groups, in which we described the purpose and procedure of the interview study, as well as eligibility and compensation. 
We also used snowball sampling ~\cite{goodman1961} by asking interviewees to share our recruitment information with their friends and colleagues, who might be interested in privacy research. 

Eligible interviewees are those who are above 18 years old and  had at least one year's experience in privacy compliance. 
Besides eligibility, we also considered the diversity of the interviewees. The recruited interviewees take different roles in privacy compliance in their companies, such as app developers, technical lead, app testers, and security engineers. The apps they work on included finance, gaming, and education. 
Additionally, interviewees' companies were located in different cities in China, with company sizes ranging from 20 to 200,000 employees. In total, we interviewed 18 participants, who spread across 10 different cities, with various roles and responsibilities in privacy compliance. 
\jingtao{We observed thematic saturation after coding 16 interviews, when no new codes were created. Therefore, we stopped interviewing new participants after 18 interviews and believe that we reached theoretical saturation.} 
\xw{Of the 18 participants, 15 were recruited through convenience sampling and 3 through snowball sampling.}
Notably, one interviewee was in Singapore, who had developed an e-commerce app for users in China, listed in Chinese app stores. We included this interviewee because they also had experience with privacy compliance in China. Interviewees' information is listed in Table~\ref{table: Participants' info}. We offered 100 RMB to each interviewee. 
The study was approved by IRB and followed the procedures of the Ethics Review Committee.

% 为了寻找实现隐私合规的参与者，我们找到了开源的隐私合规检测工具的GitHub社区，并通过社区加入到了隐私合规参与者组建的微信群。这些微信群的目的不仅仅是探讨社区中提供的隐私合规检测工具该如何使用，还为不合规的开发者提供具体的解决办法。我们通过在群内发布招募信息来邀请群内的人参加采访。招募信息中描述了本项研究的目的、过程和其中的注意事项。我们的研究还采用了滚雪球抽样的方法，推荐受访者将我们的招募信息分享给他们的好友、同事或者其他对此研究感兴趣的隐私合规参与者。
%
% 在我们的采访中，所有招募的受访者，至少是参与过实现隐私合规某一个环节，并且有着1年以上的隐私合规经验。我们周全的考虑了受访者的多样性和其开发的app的多样性。受访者包括在公司内部实现隐私合规的多种角色，如开发者、测试者、安全顾问等，而app的种类则包含金融类、游戏类、工具类、教育类等。另外，这些受访者所在公司分布在不同城市，公司规模也从最少20人左右到最大200,000人随机分布。最终我们总共采访了18名隐私合规参与者，他们分布在13个不同的城市，具有多种角色和对隐私合规具有的不同责任。值得注意的是，18个受访者中有17人来自中国，另外1人来自新加坡。这名新加坡的开发者开发了一款电商app，app的使用者有来自中国的用户，因此app有可能有上架中国应用商店的需求，故而开发者非常了解和关心来自中国的隐私合规campain。所有受访者的信息都在表1中。

\subsection{Interview Process}

We informed the interviewees about the purpose and procedures of this study and collected their consent before the interview. The interviews were conducted in Mandarin using WeChat call, lasting between 50-70 minutes. Interviews was audio recorded for transcription, with interviewees' consent. 

We first asked general questions about interviewees' industry, company size, job duties, and other background information, to collect some contextual information about their work in privacy compliance. Next, we probed their knowledge about privacy compliance in China, by asking them when, how, what and why they had learned about the privacy laws, provisions, guidelines and procedures released by the government. Based on their familiarity with privacy compliance, we probed into how they understand, interpret and perceive the privacy compliance in the development of apps, as well as how they achieve privacy compliance, such as the methods, tools, and strategies they have used. Particularly, we asked about the steps they needed to go through to have their apps reviewed by app stores and governmental agencies. We probed into the challenges, problems, and difficulties they encountered during the review. We also encouraged them to share their opinions, feelings and strategies about the review. All the translated interview questions can be found in \url{https://github.com/YkGUWbrF/SPRC}\ignore{\url{https://github.com/YJylzOyL/SPRC}\url{https://github.com/YJylzOyL/Special-Privacy-Rectification-Campaigns/blob/main/Supplementary-materia.md}}\ignore{\TODO{url}}.

\ignore{\TODO{update github}}

\ignore{\TODO{better append the translated interview questions in appendix}}

%下一步，询问受访者实现隐私合规的流程以及公司对他们实现隐私合规的影响；接着，询问受访者实现隐私合规过程中使用到的方法、工具，最后询问受访者在实现隐私合规的所有流程中可能遇到的困难和他们现在使用到的解决手段。
%
% \TODO{interview process covering all RQs}
%

% All the above information was communicated to the participants before the start of the interview.

% 在采访前，我们明确告知受访者本研究的目的和过程，所有采访内容都是匿名的，采访过程中我们也不会收集受访者个人和所在公司的身份信息。我们为受访者提供了100元人民币的补偿，大约16美元，大多数受访者都接受了这个补偿，但是仍有3名受访者明确拒绝了这个补偿，并表示愿意义务进行本研究的采访。整个研究IRB经xx大学批准，并且严格遵循XXX伦理审查委员会的程序来管理和同意数据的处理。
%
% 我们的采访采用普通话以微信音频聊天的方式来收集研究数据，所有采访时长控制在50-70分钟。采访类型采用半结构化的方式，每当我们提出一个问题，受访者都可以根据自己的经验和理解做出解答。但是采访并不是总是严格按照既定的顺序进行，而是根据受访者的回答进一步提出我们的问题。
%
% 具体来说，首先，面试官会进行自我介绍，并明确描述本次研究的背景和目的；其次，受访者会对自身情况，包括他的专业经验和公司背景信息做出简单介绍；然后，询问受访者对用户隐私有关的法律法规和政府开展的保护用户用户隐私行动的了解（如果有的话）；下一步，询问受访者实现隐私合规的流程以及公司对他们实现隐私合规的影响；接着，询问受访者实现隐私合规过程中使用到的方法、工具，最后询问受访者在实现隐私合规的所有流程中可能遇到的困难和他们现在使用到的解决手段。
%
% 如果采访过程中受访者明确拒绝回答某个问题，我们会对这个问题略过。所有采访内容我们都进行了录音for transcription，这些音频数据最后都会被销毁以消除识别出受访者身份信息的可能。所有以上内容我们都在采访正式开始前告知participants。

\subsection{Interview Analysis}

We used inductive thematic analysis \cite{braun2006using} to analyze the interviews. We first transcribed the audio-recorded interviews into text and anonymized interviewees' identifiable information. Once the transcription was done, we deleted the audio recordings to protect interviewees' privacy. Next, we read all the transcribed interviews, familiarized ourselves with the data, and independently noted down the initial codes related to interviewees' understandings, perceptions, practices, challenges and strategies in privacy compliance. These codes are meaningful labels attached to specific segments of the interview data. 
Then, we compared our initial codes with each other, went back and forth between codes and original data, discussed our interpretations about each individual code, and revised/refined the codes through multiple meetings. This step ended with compiling a comprehensive list of 875 codes. Based on the initial codes, we collated similar codes into a sub-theme, which identified 22 sub-themes. We first gathered all the original data relevant to each sub-theme, examined the codes and associated data, examined the relationships between the codes, and collapsing similar codes into a bigger and meaningful pattern. Then we further grouped similar sub-themes into an overarching theme by identifying the relationships between the sub-themes. We identified 4 overarching themes, namely privacy review workflow\ignore{Workflow of Privacy Compliance}, challenges to app developers\ignore{Challenges in Privacy Compliance}, solutions to address challenges\ignore{Strategies}, and positive changes and concerns\ignore{Current Situations}. A thematic map was thus formed, with 4 themes, 22 sub-themes and 875 codes. With the initial thematic map developed, we reviewed and refined it by checking whether the themes/subthemes captured the meanings in the coded data segments and formed a coherent pattern. %Some sub-themes were removed because there were not enough data to support them or the data were too diverse. Some themes were collapsed into each other. Some were broken down into separate themes. We also searched for data that had been missed in the earlier coding stages.

%\subsection{Final Code Book}
\section{Results}
\label{sec:results}
%In this section, we present the findings from the interviews, organized by how they address each research question. 

\subsection{RQ1: Privacy Review Workflow}
\label{subsec:workflow}

Based on the interviews, we summarize a workflow that participants regularly go through in privacy reviews. We first introduce the stakeholders in the workflow, and then report the procedures in the SPRC privacy review and app store privacy review, as two fundamental components in the workflow.

\begin{figure*}[!t]
\centering
\subfloat[SPRC privacy review workflow]{
        \label{fig: pipeline1}
		\includegraphics[scale=0.23]{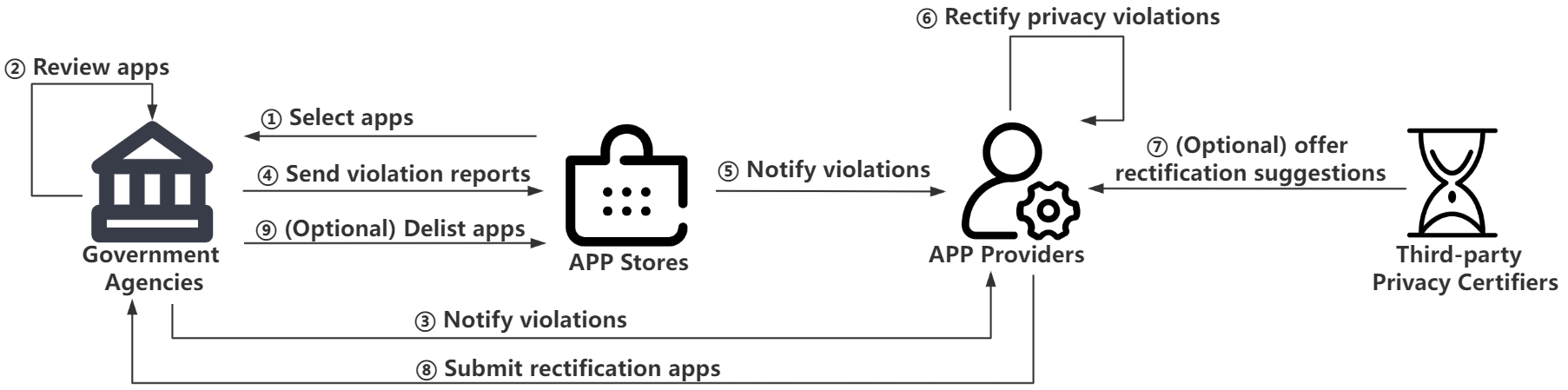}}\\
\subfloat[App store privacy review workflow]{
        \label{fig: pipeline2}
		\includegraphics[scale=0.23]{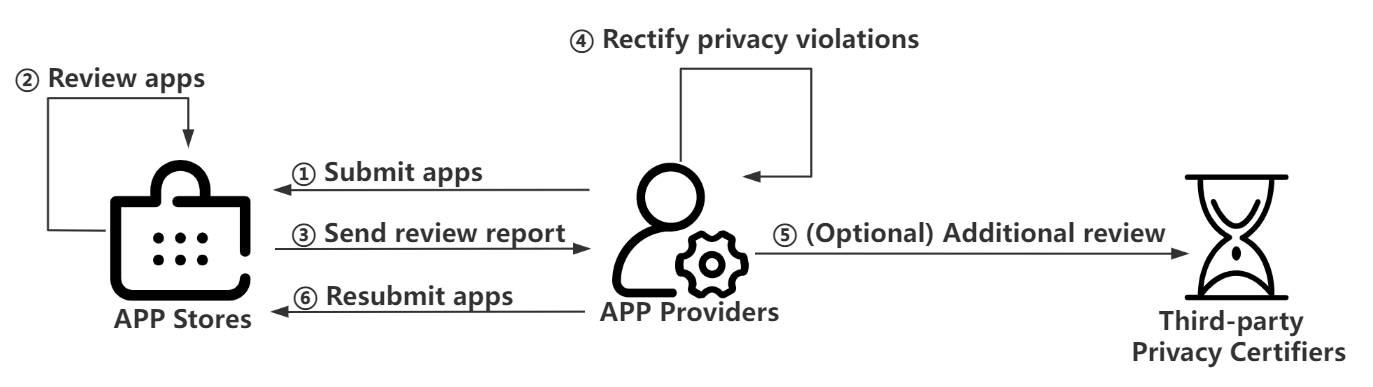}}\\
\subfloat[Information flow]{
        \label{fig: pipeline3}
		\includegraphics[scale=0.23]{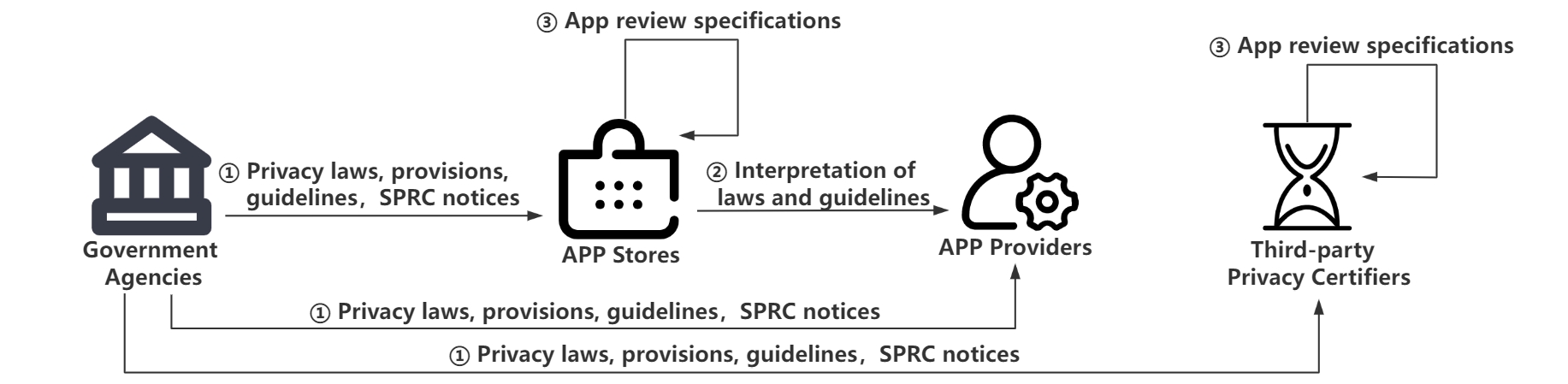}}
\caption{Privacy Enforcement Workflows}
\label{fig:pipeline} 
\end{figure*}

\subsubsection{Stakeholders}
\label{subsubsec:stakeholders}
As mentioned in Section~\ref{subsec:background_chinese}, the enforcement of Chinese privacy laws on mobile apps has primarily been driven by large-scale campaigns, i.e., SPRCs, and involves multiple stakeholders. 
In the following, we highlight the major stakeholders identified by our interviewees, and we will refer to these stakeholders throughout the paper.

\vspace{1pt}\noindent$\bullet$~\textbf{App providers} %refers to companies and individuals that develop and publish mobile apps in app stores.
refer to the companies where the interviewees are employed. %, which have varying sizes ranging from 20 to over 200K employees (as shown in Table~\ref{table: Participants' info}).
If privacy violations are detected in their apps during the reviews, app providers, such as the companies of P6, P8 and P12, will be notified by the government or the app stores.

\vspace{1pt}\noindent$\bullet$~\textbf{App engineers} are employees hired by app providers who conduct the technical design, development,  and testing of mobile apps. Most of our interviewees are in this role.
%
%For example, nine interviewees are app developers or technical leaders responsible for app development, and six interviewees are security testers or test engineers focusing on app testing. 
%
%The majority of interviewees in this study are app engineers, with the exception of P3, who served as a policy translator in the law department.

\vspace{1pt}\noindent$\bullet$~\textbf{App users} are the individuals from whom mobile apps collect personal data. In this study, interviewees also referred to app users as ``end users'', ``individuals'', or ``clients''.

\vspace{1pt}\noindent$\bullet$~\textbf{App stores} are the marketplaces where app users can find and download apps. 
Interviewees interacted with several app stores in China, such as OPPO, Huawei, and Apple Store, for app store privacy review (detailed in 5.1.2) .
%
%Note that Google Play was not mentioned since it is not accessible in China.
%

\ignore{
\begin{quote}
    \textit{``For example, channels like Huawei, AppGallery, Vivo, and OPPO. When you submit your package, they will send you an email during the submission process. They will specify where the problem occurred\ignore{, and they have a log in the form of an account reconciliation.}''} - P4
\end{quote}
% P4：比如说华为、应用宝、vivo还有OPPO这些渠道，你过包的时候他们会出一个比如说你提交过程是吧？他们会给你发邮件的，具体是哪里出问题他们会指出来的，他们有一个对账的形式的日志。
}

\vspace{1pt}\noindent$\bullet$~\textbf{Third-party privacy certifiers} emerge due to the increasing concerns and challenges in privacy reviews.
Interviewees reported that they submitted their apps to these certifiers, such as Bangcle Security~\cite{bangcle} and iJiami~\cite{iJiami}, for additional privacy review, with the aim to prevent potential privacy violations. 
Their companies needed to pay the certifiers every time their apps are reviewed. 

\vspace{1pt}\noindent$\bullet$~\textbf{Government agencies} are executive departments of China, including MIIT and CAC, along with their affiliated institutions such as CAICT (China Academy of Information and Communications Technology). We use government agencies to represent all of them. Interviewees noted that these government agencies are responsible for issuing privacy laws, provisions, and guidelines, and issuing administrative notices to launch SPRCs (detailed in 5.1.2). 

\ignore{
\begin{quote}
    \textit{``App stores perform app reviews according to some special campaigns of government agencies.''} - P8
\end{quote}
%他们都会测，他们抽查，上架的时候一般是自动化的，但是他们会抽查一些，也不知道它的规律是什么，反正有可能根据政府的一些专项行动，对某类APP做一下专项的抽查之类的。
}

\subsubsection{Privacy Enforcement Workflows}

Interviewees reported that mobile apps need to go through two types of privacy review: 
%

% 全国APP技术检测平台管理系统

%
%Second, in accordance with the requirements of SPRCs, app stores must collaborate with government agencies to conduct privacy reviews on the apps they host. This usually occurs when apps are submitted to the app stores prior to publication (i.e., \textit{app store privacy review workflow}).

\para{SPRC privacy review workflow}
As shown in Figure~\ref{fig: pipeline1}, interviewees explained that the SPRC privacy review is initiated by government agencies to select a subset of apps available in major app stores (\textcircled{\small{1}}). This review is periodical, occurring every 1-2 month. The selected subset of apps usually favors those with a large user base in China. Government agencies either collaborate with third-party privacy certifiers or build their own certification services\ignore{~\cite{}}, to conduct this periodical privacy review (\textcircled{\small{2}}). For instance, P3 and P8 told us:

\begin{quote}
    \textit{``From what I know, the MIIT has entrusted the CAICT to build their own [app privacy review] platform.''} %- P3
\end{quote}
% 我知道的是像工信部它是委托信通院去做支撑，自建了一个平台，然后结合自动化的技术和人工的技术，他们的数据覆盖范围是比较大的 - p3

\begin{quote}
    \textit{``The MIIT also invites some third-party privacy certifiers.''} %- P8
\end{quote}
% 工信部也会组织一些第三方的评估机构，他们去查，通报的时候好多已经市场都没有报过，他们会报出来。- p8
%
% P3: 中央部委做检测它不是兜底，它是抽查还是监督
% P8: 一方面其实就是说应用太多了，公司太多了，光靠以前前期的及时抽查，前期用户量大的比可以也会优先，然后头部的一些APP整改一遍，然后再去，但是太多了，几十万款APP可能是抽查，但是也不能覆盖所有的，就需组织一些培训法律法规，让大家去学习自查自纠，不可能完全靠通报整改的一个途径。
%
%Then, the government agencies conduct privacy reviews on these apps (either using their own services or third-party services) to identify privacy violations within the apps . 
%
%
If an app is found to have privacy violations, the government agencies will issue a public bulletin regarding the app's privacy violations on government websites (\textcircled{\small{3}}).
Subsequently, the government agencies will send the app review reports containing the violations to the app stores (\textcircled{\small{4}}).
App providers are then notified of the privacy violations of their apps and receive the reports from the app stores (\textcircled{\small{5}}).
The app providers need to address the privacy violations based on the reports (\textcircled{\small{6}}).
If the app providers have purchased the service from the third-party privacy certifiers, the third-party certifiers will provide compliance services and rectification suggestions to app providers based on violation reports (\textcircled{\small{7}})
It's important to note that, due to the presence of privacy violations, app providers are only given a short period of time, typically 5 business days, to rectify their apps.
Once the app providers address the violations in a new version, they will submit it to the government agencies for a second review (\textcircled{\small{8}}).
If the new version fails to address the violations adequately, government agencies will notify app stores to remove the apps from their listings (\textcircled{\small{9}}).
For example, P3 noted:

\begin{quote}
    \textit{``They [government agencies] implemented two mechanisms, notification [placing on bulletins] and app store removal. If the rectification is not completed within five days, then the app is removed from the app stores.''}% - P3
\end{quote}

\para{App store privacy review workflow}
\ignore{As reported by the interviewees} As shown in Figure~\ref{fig: pipeline2}, the interviewees reported that once an app is developed, its provider submits it to app stores for privacy review before making it available for end users (\textcircled{\small{1}}). The app stores review the app in accordance with their app review specifications (\textcircled{\small{2}}), which are drafted based on the privacy laws, provisions and guidelines from the government agencies. If no privacy violations are identified during the review process, the app will be published in the app stores without any changes. However, if privacy violations are detected, the app stores will generate a privacy review report and send it to the app provider (\textcircled{\small{3}}). The app provider must address the violations in the review reports (\textcircled{\small{6}}). It's worth noting that this rectification and resubmission process may occur multiple times until all violations are appropriately resolved.
For app providers are concerned with potential privacy noncompliance, they may choose to conduct an internal privacy review on their own (\textcircled{\small{4}}) or pay third-party privacy certifiers for pre-review (\textcircled{\small{5}}), before submitting to app stores. These reviews (\textcircled{\small{4}} and \textcircled{\small{5}}) are optional and can take place in any sequence. For instance, P9 shared:

\begin{quote}
    \textit{``We need to publish our apps on five different app stores, Xiaomi, Huawei, and then Vivo, and AppGallery. When you submit your app, there are usually one or two app stores that don't pass the review, and then you have to revise.''} %- P9
\end{quote}
% app provider整改完后提交应用商店上架 - P9
% P9：是的，包括国内这些应用市场，基本上因为我们是要上架5个应用市场的，小米、华为，然后vivo、应用宝这几种，你上架指定是有几个是审核不通过的，然后你再根据他的驳回建议，然后你基本上你你把他提的建议改完之后基本就算过了，也不会再给你提新的问题，就是每一个版本基本上架都会有这种问题。就上架一轮，然后有一两个驳回了，然后就改。

% code and quotes
% procedure ③
% 应用商店反馈回整改的问题后，app provider的解决方式 - P11
    % P11：（Q：应用商店它返回检测不合规的结果之后您公司的处理的流程是什么？）先安全或者法务部门，首先要确认这个信息。给到安全部门，确认这个问题是不是真的存在。有时候上面的信息不一定准确。我们需要核对的。所以说有时候这个东西它不是很稳定。... 如果存在的话， 你和法务去确定这个场景是否必须的，如果是必须的那就加上去（再向其他办法解决），如果不是必须的，那就去改。首先是我过来我们这边去核对这个信息是否真实存在。因为我们这边的话它在上架之前是做过检测的，如果没有检测出，那说明这个东西它大概率是不存在的。如果复现如果真的存在，说明前期检测没有覆盖到位，如果不存在，那就厂商这边误报重新提交。
% 应用商店反馈给app provider的信息 - P4
    % P4：比如说华为、应用宝、vivo还有OPPO这些渠道，你过包的时候他们会出一个比如说你提交过程是吧？他们会给你发邮件的，具体是哪里出问题他们会指出来的，他们有一个对账的形式的日志。

% procedure ④ and ⑤
% 合规整改需要大量人力物力，同时也可能交给第三方检测机构来整改	P7
    % P7：这种问题的成本相当大，你有这个问题你就要加班嘛。 ... 但是考虑到有一些公司可能不具备检测的实力，很多的时候把检测的就交给第三方。
% 应用开发商会购买第三方安全服务。这种服务更专业，具体来说，会打包提供其他安全服务，如加固；第三方服务有客服售后对应用厂商做支持保障	P18
    % P18：目前还是通过第三方检测，我们一般不关心他们，也不会去直接去读它，国家变化了，我们立马去变，主要还是按照第三方的，他们更新我们就跟着更新，那么检测出来的问题我们就去更新，（明白了。）所以我就跟他们搞。第三方收费的爱加密，刚才你说的爱加密之前还有什么？百度应该都有一些，之前应该都是厂商他们自己购买了一些服务，我们用他们服务就行了。还有阿里的EMAS架构也有。... 收费的更专业一些，而且有人维护和更新，有（隐私合规）问题，你可以找到他

% procedure ⑥
% app provider整改完后提交应用商店上架 - P9
    % P9：是的，包括国内这些应用市场，基本上因为我们是要上架5个应用市场的，小米、华为，然后vivo、应用宝这几种，你上架指定是有几个是审核不通过的，然后你再根据他的驳回建议，然后你基本上你你把他提的建议改完之后基本就算过了，也不会再给你提新的问题，就是每一个版本基本上架都会有这种问题。就上架一轮，然后有一两个驳回了，然后就改。

\para{Information flow}
In addition to the aforementioned workflows, interviewees also mentioned an information flow stemming from privacy laws (as depicted in Figure~\ref{fig: pipeline3}).
Specifically, app providers, app stores, and third-party certifiers take the privacy laws, provisions, and guidelines issued by government agencies as input (\textcircled{\small{1}}). 
Subsequently, app providers strive to develop privacy-compliant apps based on their interpretation of the laws, provisions and guidelines (\textcircled{\small{2}}).
App stores and third-party certifiers build their app review specifications in line with the administrative notices of SPRCs, based on privacy laws, provisions, and guidelines. Interviewees, such as P5 and P18, indicated that the app review specifications would update frequently with the updates of SPRCs (\textcircled{\small{3}}).

\vspace{0.5em}
\noindent\fbox{%
    \parbox{0.95\columnwidth}{%
    \textit{\xwnew{Interviewees report that }SPRCs involve various stakeholders such as app providers, app stores, third-party privacy certifiers, and government agencies. In SPRCs, mobile apps undergo two types of privacy reviews: the SPRC privacy review and the app store privacy review.}    
    }
}
\vspace{0.5em}

\subsection{RQ2: Challenges to App Developers}
\label{subsec:challenges}

We report three major challenges that interviewees have encountered in the privacy reviews and privacy compliance.

% conclusion -> what -> why -> what result
% 1) 两个不同开发者对隐私合规的实现不一致，原因是对同一份notice的理解有偏差，以及需要根据自己app功能、业务实际需求来满足合规（可能不需要这条）
% 2) 同一个app在同一个应用商店的不同时间点检测结果不一致，（interviewee推测）是因为应用商店的检测工具的变化导致的这种现象 (可能要放在time sensitive里)
% \TODO{Time sensitive}
% 3) 同一个app在不同应用商店的检测结果不一致，不一致的原因有A）对隐私合规的理解不一致（root cause？）、B)使用的检测工具不一致、C)应用的检测环境（smartphone）不一致、4) 同一个应用在不同应用商店的检测结果的表现形式不一致，（这个好像没有原因）
% D)应用检测的严格程度不一致
% device as an example
% 5) developer认为的个人敏感信息和应用商店检测的不一致
% 6) developer需求的SDK的配置和SDK提供的不一致

\subsubsection{Inconsistencies}
\label{subsec:inconsistencies}

\para{Same app receives inconsistent privacy review reports from different app stores}
Eight interviewees (\xw{from both small companies, such as those of P7 and P9, and relatively large companies with over 1,000 employees, such as those of P1, P8, P14, P16, P17, and P18}) mentioned that they received privacy review reports with inconsistent results when submitting their apps to multiple app stores for review\ignore{ \jingtao{no mater how big the company is}}. 
Each app store develops their own privacy review specifications and tools based on the laws, provisions and guidelines issued by the government. 
They may interpret and implement the laws differently.
The interviewees attributed the inconsistency of the results to the different detection tools adopted by the app stores, as app stores do not always update their tools at the same time.
%
%App stores usually developed their own tools for detecting privacy violations and update those tools according to the privacy guidelines and the changes of SPRCs from government agencies. 
%
They also do not implement all the updates at once, but selectively deploy important updates that they value first. 
\xw{P9 pointed out that some app stores, citing Xiaomi as an example, } even outsource their privacy reviews to different third-party privacy certifiers, each providing different services or offerings.  
These issues led to privacy violations reported by one app store but not by the other.
As a result, app developers need to invest a large amount of time to triage the reports from the app stores and explore the reasons behind the inconsistent detection results, which wastes their time that were supposed to be spent on privacy violation mitigation.
For example, P9 noted:
% 是不一致的，他们每个公司都有自己的一套检测的一个工具，但是他们把一个侧重点，因为他们也是要更新迭代的，就是每个公司的侧重点它也不太一样，但是大致是差不多的，大致都是跟政策走。
%
\begin{quote}
    \textit{``%You know the vetting of app store is not that strict, right? 
    Those app stores often outsource their privacy vetting to over ten different third-party companies, and therefore, the results sometimes are not consistent.''} %- P9
\end{quote}
%因为上架应用市场一直管的不是很严，知道吧？他们因为这些个应用市场都会和十几家的第三方外包公司，然后去提升他们的检测工具。是他们也会去外包给其他的公司去检测，这个东西，然后所以说有的时候不是很一致，然后上架应用市场管的也不是特别的严，也相对于我们自己和检测公司去检测的话，要差好多，因为我们自己找第三方，有的公司只要是有一点风险的话，他都会给你提出来。
% (检测结果)是不一致的，他们每个公司都有自己的一套检测的一个工具，但是他们把一个侧重点，因为他们也是要更新迭代的，就是每个公司的侧重点它也不太一样，但是大致是差不多的，大致都是跟政策走。

In addition to the inconsistent app review results from app stores, the quality of the app stores' reports, such as the granularity and comprehensiveness of results, also varies between app stores.
Participants reported that some app stores, such as OPPO and Vivo, present detailed information about privacy violations in their reports, including class/method names and app stack traces, making it easy for app developers to pinpoint the exact reason behind the violations.
However, other app stores, such as Xiao Mi, failed to provide these details. They simply listed the general privacy violations without why and where, leaving the developers to explore on their own.
Such exploration demands significant effort from developers, since searching for the causes (e.g., piece of code) of the violations and confirming them can be tedious and time-consuming. It becomes even more difficult considering that the code that leads to the violations may not be developed by their own teams, and in many cases, not even from the same company (e.g., third-party code). For example, P4 told us:

\begin{quote}
    \textit{``For instance, from my perspective, OPPO and vivo provide similar reports in an Excel spreadsheet, detailing stack information. In contrast, platforms like Xiaomi and Yingyongbao don't provide such details, just a table, very general about the problems, leaving you to investigate yourself. This can be rather troublesome.''} %- P4
\end{quote}
% P4: 肯定的，因为他们都有自己的检测部门，像比如说在我看来就OPPO跟vivo他们检测的报告是类似的，出一个Excel表格，然后告诉你堆栈的信息，像比如说小米应用宝这些就不会那么详细，它会他可能那边后台会有一个列表，很笼统的告诉你这个规则是哪里出问题的，但是具体细节他不告诉你，然后让你去查的，这就是比较头疼的一个问题。

% 2） 开发者和应用商店运行app的手机不一样。
\para{App developers' self-testing results are different from app stores' privacy review reports}
The privacy violations identified by app stores often do not show up when app developers test their apps themselves. %\ignore{The privacy violations identified by app stores often differ from the actual app behaviors observed by app developers.}
Two interviewees (P7, P14) noted that app stores reported violations in their apps for collecting sensitive data and for posting permission requests before users accept privacy policies. But these violations could only be found in the devices used by the app stores. 
When the app developers tested the apps on their own devices, they had a difficult and frustrating time reproducing the reported violations because they were unable to observe the behaviors in the apps.
The major cause of this issue is that the app stores used different devices to test the apps from the app developers, and the app violating behaviors show up only on those devices. For example, P7's app turned out to collect additional data on app store's device (Huawei Mate 50 series devices). P14's app tended to post additional permission requests before privacy policies on app store's device. But these issues did not show up in app developer's devices. It is impractical to ask app developers to adopt the same devices as the app stores', because they are unaware of the devices that the app stores would use, among hundreds of different existing devices, before the review. For instance, P7 said:

\begin{quote}
    \textit{``The device model, which can sometimes lead to different test results... testing on the Huawei Mate 50 series (in app store review), an app was found to collect data... However, this issue was not observed in other device models.''}%- P7
\end{quote}
% P7: 大部分问题可能都是一样的是的。主要还有一个点有点难说的，就是它的测试机型，有的时候会导致测试结果的不同。(机型会影响检测的结果吗？) 对华为最新的最新的mate50系列。上次就检测出来，检测的时候，它会有一个第三方SDK在同意隐私政策前就做条件获取的行为。但是在其他机型上是没有这个问题的。

%\begin{quote}
 %   \textit{``Because after the updates of Huawei..., certain devices will directly ask for permissions for collecting app package names.''} - P14
%\end{quote}
% P14: 当然是应该弹出来的，但是在这个位置的话，我们当时是没有匹配到人家的机型。是的，因为华为这边做它系统升级之后，可能就是在机型在它这个情况当中，它信息就之前我们收集应用列表信息的话，它是不会作为权限来直接弹出的，但是在特定一些机型上边，可能就是后续就是可能权限会成为像定位电话这一类通用权限它直接会弹出来，就会系统的就是把这个系统内部会做一个权限申请。

% Developer认为的个人敏感信息和应用商店检测的不一致
\para{App developers and app stores define sensitive personal data differently}
App developers and app stores often have an inconsistent understanding of what personal data should be considered sensitive and warrant better protection through the enforcement of privacy laws.
% 
%\ignore{\TODO{why this inconsistent understanding? }}
This inconsistency mainly concerns user-specific and device-specific data. Particularly, P9 believes that data directly associated with users (e.g., phone number, family information) is highly sensitive, while device-specific data alone (e.g., Android ID, IMEI) is not considered sensitive.
However, app stores and the tools they use focus on detecting the unauthorized collection of device-specific data, providing almost no coverage for user-specific data.
This issue is primarily caused by the limited capabilities of their tools to identify user-specific data. App stores' tools can easily detect device-specific data by monitoring a fixed set of system-level APIs that emit the data. But their tools are not capable of identifying user-specific data due to the contextual nature of the data to each app. For example, P9 mentioned:

\begin{quote}
    \textit{``In fact, for us [a business-to-business (B2B) app], user data such as national ID, passports, phone numbers, and family info are considered sensitive data... But Android ID, Mac addresses, IMEI are not.''} %- P9
\end{quote}
\para{Popular apps are reviewed more strictly than unpopular apps}
Interviewees reported that different apps can undergo different levels of privacy review by app stores.
For most unpopular apps, app stores usually run automated analyses to detect their privacy violations. However, they apply additional manual analysis to popular apps, such as DiDi (a leading taxi-hailing app) and Meituan (the most used food-delivery app in China), which are used by nearly everyone in the country. One reason behind the different reviews is that the government-level privacy review (SPRC privacy review) tend to focus on popular apps with a large user base in China. Hence, app stores follow this pattern and perform extra manual analysis on popular apps. For instance, P5 said:

\begin{quote}
    \textit{``For less well-known apps, the review process is primarily automated... There can be exceptions for particularly famous applications with a wide user base, such as Didi, Meituan, or Douyin. These popular apps often undergo some extra manual review.''} %- P5
\end{quote}
% 反正因为就是说应用商店，因为现在国内好像几百万个APP。他不可能做到全部的人工，他也是用机器自动，它的一些结果都是自动化，它不会存在，可能也会有部分特别出名的一个应用。就像这种用户特别广，你比如像滴滴、美团这种全国人民全都在用，还有抖音，像这种的会加入一些人工的审核，像一些不太知名的APP，基本上都是用自动化技术的手段去审核。

Automated analysis is more efficient, but often fails to reveal privacy violations hidden in deeper program paths (e.g., requiring more user interactions to trigger), whereas manual analysis can complement the detection of these violations by exploring deep paths. 
Thus, unpopular apps are not evaluated as thoroughly as those popular apps, leaving questions about the extent of privacy assurance for these unpopular apps, which also collect personal information from end users. 

% 5） SDK和app的需求不一样。
\para{The functionalities offered by third-party SDKs do not align with the privacy review requirements}
Third-party SDKs play a crucial role in the supply chain of almost every app. When integrating these SDKs, interviewees anticipated incorporating only the functionalities required by their apps to meet the privacy compliance requirement in the privacy reviews.
However, interviewees reported that SDKs often introduce unnecessary functionalities. For instance, P11 described a situation where their app initially needed an SDK solely for displaying maps based on user location data. Nevertheless, P11 ended up integrating a multi-functional SDK that not only displayed maps but also included voice recording features.
% 这里SDK获取的语音信息和app本身的业务无关，所以认定为超范围收集用户信息，也就违反了隐私合规，而这并不是app开发者的本意。
%
The issue arises because the SDK must be used all-or-nothing, preventing P11 from selectively enabling only the necessary features. As a result, this led to the collection of unnecessary personal data, which is deemed redundant and a violation in the privacy review.
This challenge reflects the limitation of third-party SDKs and the challenges posed to privacy compliance, as they lack the flexibility to be configured and customized to the specific needs of apps:

% SDK问题凸显凸显出来，并且变得严重

\begin{quote}
    \textit{``For example, your app requires location access functionality, while other apps may need recording or other features. Considering the variety of apps, they only offer one SDK, but it's used in different scenarios in different apps... For third-party SDKs, it's not feasible for them to customize an SDK specifically for your app's needs.''} %- P11 
\end{quote}

\subsubsection{Yesterday's Compliance, Today's Noncompliance}

%\xw{made slight changes}
% 1)privacy rules的更新/变化导致开发者实现隐私合规的变化和隐私合规检测工具的变化
\para{Frequent updates to app review specifications make previously compliant apps noncompliant}
\ignore{Privacy laws, provisions, and guidelines, typically issued by government agencies, tend to remain relatively stable without frequent changes. However, as part of the SPRCs, government agencies regularly issue administrative notices to key stakeholders, particularly app stores, to address gaps in privacy law enforcement.
% 工业和信息化部关于进一步提升移动互联网应用服务能力的通知
An example of the notices is the MIIT Notice [2023] No. 26, titled ``Notice on Further Enhancement of Mobile Internet Application Service Capability''~\cite{miit2023no26}.
In response to such notices, app stores are required to update their app review specifications, such as expanding the requirements for sensitive data or adjusting the intensity of reviews.
% https://ythxxfb.miit.gov.cn/ythzxfwpt/hlwmh/tzgg/xzxk/dxhhlwyw/art/2023/art_5ae5dfa171d2407d9178ecda973c9817.html
%
}
The frequent changes of app review specifications of app stores have posed extra challenges to app developers. Five interviewees (P1, P3, P5, P14, P18) reported experiences with frequent changes in their apps' compliance status. In other words, an app initially identified by app stores as privacy-compliant can swiftly become non-compliant due to the updates in the app review specifications. Consequently, the app must undergo an unexpected rectification process, potentially requiring re-submission to app stores. This incurs a significant cost in terms of resources, money, and time to the app's provider, like what P5 said:

%\ignore{Privacy laws, regulations, and provisions, once enacted, tend to remain relatively stable without frequent changes.
%
%However, as part of the SPRCs, government agencies frequently release new privacy enforcement guidelines to stakeholders, particularly app stores, to enhance the enforcement of privacy laws. 
%
%For example, the China Academy of Information and Communications Technology (CAICT), a research institute affiliated with the MIIT, often releases white papers~\cite{} and online educations materials~\cite{} with new guidelines.  
% http://www.caict.ac.cn/english/research/whitepapers/202112/t20211224_394504.html
%
%In response to these guidelines, app stores need to add new rules to their tools, such as extending the coverage of sensitive data, or adjusting the intensity (or levels) of enforcement, e.g., warning, fines or app delisting\TODO{Better examples may be the changes in the government's notice. And need address that the adding behavior is to satisfy privacy compliance}. 
%
%These newly added rules have been found to cause issues for app developers. For example, five interviewees (P1, P3, P5, P14, P18) reported experiencing frequent changes in their apps' compliance status.
%
%That is, an app identified as privacy-compliance can rapidly become non-compliant due to the introduction of the new rules. 
%
%As a result, the app has to undergo an unexpected rectification process, which may involve re-submission to app stores. This incurs a significant cost in terms of resources, finance, and time to the app's provider.}

\begin{quote}
    \textit{``There was a time when Bluetooth information was not considered personal data, but within a month, it was reclassified as such, which put us in a difficult situation, because we have to adapt accordingly''} %- P5
\end{quote}
% 然后还有其实就是说监管变化也比较频繁，比如说反正因为现在比如说前段时间，蓝牙信息它不是个人信息，但是可能这过了一个月蓝牙信息又是了，这就搞得很被动。其实还是是因为他们一变，我们这一些做安全的你不要跟着去变，因为所以说这块会有一个延迟。就是说如果频繁的变的话，其实我们这块比较模糊。

\ignore{\begin{quote}
    \textit{``Of course, the standards we set are based on national regulations issued by the Ministry of Industry and Information Technology (MIIT). This applies to both third-party organizations and the app stores where app is listed. Moreover, these standards are not static; they can change over time. For instance, earlier, there might not have been a requirement for permission collection, but later, as requirements evolve, we incorporate these changes. Additionally, apps themselves are continually being upgraded to comply with new standards.''} - P18
\end{quote}
}
% don't quite understand what is "权限收集"
% 当然我们定的一些都是按照国家规定，工信部来的，是不管第三方机构还是应用上架一些市场都是以那种作为依据，而且也不是不变的，可能你很早的之前权限收集，他没有要求，后面他可能有要求变化的，有要求的话我们就把它加进去。对新的，而且APP有些东西也是一直升级。 - P18

Complicating the problem is the use of open-source tools.
App developers often want to check their apps using open-source tools before submitting to privacy reviews.
However, these tools are not always updated to catch up with the changes in the privacy review specifications.
As a result, an app that these tools reported as compliant can have violations in the privacy reviews. For example, P11 said:

\begin{quote}
    \textit{``%Sometimes, the need to update [open source] tools can also be a factor. 
    Many issues arise from the tools not being updated promptly, then submissions were returned after review.''} %- P11
\end{quote}

%\ignore{\begin{quote}
    %\textit{``Sometimes, the need to update [open source] tools can also be a factor. Many issues arise from the tools not being updated promptly, leading to submissions being returned after review. However, under normal circumstances, this kind of return due to outdated tools is not a common occurrence.''} - P11
%\end{quote}}
% 有时候(开源)工具像有时候工具也需要去更新的，很多问题存在于就是说可能工具更新不及时导致的，然后提交之后被返回，一般情况下不存在这种被返回的场景。 - P11

% A）政府政策会更新/变化
% 政府政策规范是所有开发商、应用商店的检测标准来源。但这个检测标准是不断更新的。		P18
% 政府监管标准是在动态不断更新的		P3
% 不同时期的整改流程是有变化的	P1
% 从业者需要不断获取新的合规消息和知识，因为技术在更新，政策在更新	P14
% 政府法律法规中有关隐私合规的标准变化比较频繁导致合规困难	P5
% B）工具会根据政策的更新而更新
% 不同应用商店的检测标准和工具更新时间都不一致	P18
% 应用开发商使用自研的检测工具，优点是更新更快覆盖更广。而开源工具覆盖不到的地方很多是因为更新不及时导致的	P11

% 
% 2)由于app更新的功能/模块导致需要频繁对app进行合规检测（自测/应用商店检测）
%\xw{made slight changes}
\para{Constant updates to the apps make long-time privacy compliance unattainable}
App providers need to regularly update their apps to keep up with technological advances and enhance the user experience. However, these updates can potentially introduce privacy violations to the apps.
Interviewees noted that many app updates are delivered through dynamic features that don't require the apps to be resubmitted to app stores. This creates challenges for privacy enforcement from the perspectives of both app stores and app providers.

Currently, app stores only review the snapshot taken during the initial app submission and lack practical ways to monitor and review dynamic updates for privacy concerns. As a result, an app that was compliant at the time of submission may no longer be compliant due to subsequent updates.
App stores, identified by SPRCs as partially responsible for the apps they distribute, cannot provide a reliable guarantee to app users about the privacy status of apps, even those they have reviewed.
For instance, P18 mentioned that:

\begin{quote}
    \textit{``Because those updates are not approved by them [app stores], nor will there be reviews... they will not know that the app has been updated''} %- P18
\end{quote}

% 00:28:11 说话者2:
%但是有一些大厂可能在用，一个是比如你苹果的，因为它要求比较严格，不允许你自动去更新。
%因为你这种更新是没有经过他们同意的，也没有检测的，（是的）。后又偷偷把它更新掉了，因为更新的东西他也不知道对吧？是国内的话可能比较松一点，但是后面这个趋势应该都不按你更新报，包含升级安装，你在应用那里面升级，它后面估计应该都会不容许尽量去在应用市场去把它更新，跳到应用市场，对吧？就像你说的自己内部某个模块更新，这种应该是类似的。明白。- P18

Second, these app updates also pose challenges for app providers striving for privacy compliance.
Three interviewees (P6, P9, P12) pointed out that their apps have such updates, and thus it is infeasible to maintain privacy compliance all the time.
In particular, P12 explained that, as part of a security testing team, conducting privacy reviews for these app updates was infeasible due to the sheer volume of items to test. Due to this, P12 believes that achieving complete privacy compliance is not possible:

\begin{quote}
    \textit{``However, we can't guarantee 100\% compliance since there are too many items to test. An app might be compliant at the time of testing, but later additions of new features or modules could result in non-compliance. Therefore, compliance is time-bounded, or it can only be approximated by removing major issues, but there is no absolute and long-term privacy compliance.''} %- P12
\end{quote}

\subsubsection{Lack of Institutional Support from App Providers}
\label{subsubsec:insufficient}
%
%Our interviewees cover multiple app engineer roles within app providers, such as app developers and technical lead, app testers, and security engineers. 
%
%These engineers often express a sense of lacking support in achieving privacy compliance.

\para{Lack of support for studying privacy laws}
%
%\ignore{Achieving privacy compliance requires a deep understanding of privacy laws and regulations. Ideally, the analysis of these laws should involve legal professionals.
%
%However, four interviewees (P8, P9, and P10 in an app developer role, and P15 in a security engineer role) reported that they have to spend a significant amount of effort studying the laws by themselves and tracking changes of the laws.
%
%For example, P10 complained that law terms are complex, and there should have been some legal professionals to help them to figure out specific terms that apply to them and piecing these terms together to form actionable items for engineers. However, the app providers tried to minimize costs by not hiring a legal professional and asked app developers to take on the responsibility of translating the laws.}
%
Achieving privacy compliance requires a deep understanding of privacy laws, provisions, and periodical SPRC-related administrative notices issued by the government agencies\ignore{, as well as the frequent updates of app store priacy review specifications}. 
%in the periodical government-level SPRC privacy reviews and app store privacy reviews 
Whenever the government agencies issue an administrative notice for SPRCs, the review specifications will be updated, and the app developers are required to update their understandings as well. 
Ideally, these legal and governmental documents should be translated by legal professionals first before consumed by app developers.
Nevertheless, four interviewees (P8, P9, and P10 in app developer roles, and P15 in a security engineer role) reported that they had to invest a significant amount of effort studying and tracking changes in the laws by themselves.
\jingtao{These interviewees work at companies of different sizes: P8's company has 5,000+ employees, while P9's and P15's companies have 20+ and 100+ employees, respectively. This indicates that both large and small companies lack support for studying privacy laws.}
For instance, P10 complained about the complexity of legal terms, and highlighted that there should have been some legal professionals to help them decipher specific terms applicable to them and piecing them together into actionable items for engineers. However, the app providers attempted to minimize costs by not hiring a legal professional and asked app developers to take on the responsibility of studying and interpreting the laws.
Additionally, participants reported that statements are lengthy and designed to cover a broad range of mobile apps. As the app developer for a specific app, they must read all the statements, figure out the specific statements applicable to their app, and translate the statements into concrete technical requirements based on his own understanding.
Lacking professional legal and privacy support not only increases burdens for app developers but can also lead to erroneous or biased understanding since most engineers do not have expertise in laws. For instance, P10 said that:

\begin{quote}
    \textit{``the documents are often lengthy and complex... [There is a need] to distill some useful points for developers like us. Simplify them, then list them one by one, so we can check against these points. Translate them into actionable items that developers can easily implement. Identify which legal clauses and regulations we need to adhere to... it requires piecing several parts of laws together.''} %- P10
\end{quote}
%P10: 没有(解读过具体法律条文)，（因为）都是长篇大论的， ... （但需要）总结出一些对我们这种开发者比较有用的一些东西。精简化，然后一条一条的列出来，我们再根据这些东西你可以去检查一下。 ... 你要把它转译成开发者会比较容易去做的东西。有哪几个是我们现在要执行的条款法律条规。一般人没有办法一次性全部说完，好几个，然后它又不是一个很完整的东西，你得好几个拼在一起

% 从业者对隐私合规政策的关注需要大量精力和成本		P8
% 开发者对隐私政策法律法规没有详细了解，内容太长	P10
% 公司内部缺乏沟通，开发者需要时间成本来熟悉使用检测工具复测提出的隐私合规问题，并在整改好后证明问题确实已经被成功解决。	P9
% 开发者认为政策需要专业人士去解读	P8
% 开发者在自己学习法律条款时会遇到难以理解的条款   	P15

\para{Lack of resources to self-validate privacy compliance}
Participants complained that they had to develop tools on their own to self-validate privacy compliance before the government and app store privacy reviews. 
As the cost of failing the privacy review is huge, such as being publicly notified and unlisted from the app stores, app providers hope to self-detect potential privacy violations in their apps before officially submitting for app stores' or government's review. 
The self-detection requires a testing environment and a set of privacy violation detection tools, similar to PrivacySentry~\cite{privacysentry}, and Google Checks~\cite{checks}.
% https://github.com/allenymt/PrivacySentry
% https://checks.google.com/
%
App developers were expected by their institutions to develop these tools on their own if they want to self-detect their apps. 
However, app developers mostly specialize in implementing core app functionalities. They may not know how to build a testing environment, or deploy tools to assess an app's compliance with privacy regulations.
They lack dedicated institutional resources, such as tools and privacy testers, to help them validate their apps' compliance status.
As a result, app developers, after implementing the main functionalities, are further tasked with proving that the functionalities are privacy-compliant. They are either compelled to rely on open-source privacy violation detection tools or to learn the skill set and implement their own tools.
The former option provides no guarantee, as open-source tools are from third-party and unknown entities, while the latter incurs significant learning curves and implementation efforts.
Hence, having institutional resources (e.g., tools and dedicated testers) in place would significantly reduce the burden on app developers, and further simplify and enhance privacy compliance. 
For example, P9 mentioned that:

\begin{quote}
    \textit{``For developers, ... the focus is typically on their own field, with only superficial knowledge about other domains, like basic understanding at best. For instance, tasks like packet sniffing are generally within the skill set of most programmers. However, developing a tool for detecting (privacy compliance) issues is not quite feasible, as this pertains to specialized tasks in the field of security.''} %- P9
\end{quote}
%P9: 但是对于开发来说，... 会专注自己领域，一般其他的领域也是接触不深，稍微有一些了解，你要是说一些的抓包等基本上程序员都会是吧？但是你要说开发个检测工具检测（隐私合规），就不太（可行），这是做安全方面的一些东西。

%\begin{quote}
 %   \textit{``there isn't a dedicated position for this task. Some companies do require such a position, which falls under compliance management.''} %- P8
%\end{quote}
% P8: 实际上没有专门的人去跟这些， ... 没有一个专门的岗位干这个。有的公司是需要有这个岗位，属于合规管的岗位。因为它其实也涉及到不光是开发，其实也涉及到产品或者技术，包括测试，可能还有信安部门等等，都是需要去考虑隐私合规的相关的一些法律政策。

\para{Lack of support for effective communication}
As the privacy reviews involve multiple stakeholders, such as app stores, third-party certifiers, government agencies, and app providers, interviewees (P3, P7, P11, P17, P18) highlighted the need to perform additional communication duties with these stakeholders about privacy review results, which \jingtao{happened to both small (e.g., P7) and relatively large companies (e.g., P3, P11, P17 and P18)}.
\xwnew{The interviewees noted that these duties create a heavy burden beyond their regular job responsibilities, resulting in difficulties to address privacy violations within the allowed timeframe (i.e., 5 business days).}
%
%The communication occurs between different parties.

When an app is flagged for privacy violations, %developers are tasked with driving and managing a long communication process that bridges internal and external stakeholders.
the developers first initiate internal discussions, by reaching out to legal and security teams to clarify the reports and determine whether they agree that a violation has indeed occurred.
In the case of disagreement, the developers are then expected to convey the perspectives of the internal teams to app stores, third-party privacy certifiers or government agencies.
After that, developers may need to persuade the reviewers that the reported violations are false positive.
The aforementioned communication process is followed by additional communication required by technically reproducing the identified violations.
For example, P11 mentioned that:

\begin{quote}
    \textit{``First confirm the message [privacy violation reports] with legal teams... then send to security teams to confirm the reports... then confirm with legal team whether this [the feature that causes violation] is necessary... if we can not reproduce, we ask vendors [app stores] to reproduce. %They should be able to since the results are from their reports.
    ''} %- P11
\end{quote}

When privacy violations are caused by third-party SDKs (e.g., for P7), developers need to communicate the reported violations with the SDK developers and request a patched SDK. Unfortunately, this often turns out to be a challenging task that requires many rounds of communication, as many SDK providers are not responsive or not reluctant to patch their SDKs based on individual developer's requests.

In addition, interviewees reported that communicating as contractors about privacy compliance is most challenging. Some app developers worked as a contractor that develops apps for another company. 
When the app fails the privacy review due to noncompliance, the contractor developer is expected to explain the privacy violations to the company. 
However, the company is more concerned with the main functionalities of the app, and less interested in privacy compliance. There are often situations where functionalities conflict with privacy compliance (e.g., resulting in noncompliant data collection). To resolve the issue, developers need to engage in intensive communication to help the company understand the importance of complying with privacy regulations. For example, P17 told us:

%\ignore{Having a specialist in charge of this communication will free developers from these efforts and allow them to focus more on implementing privacy-compliant features.}

\begin{quote}
    \textit{``The biggest challenge lies in the requirement confirmation phase because sometimes clients focus more on functionality and may not care about privacy compliance, even though these must be addressed. So communication on requirements can incur significant costs.\ignore{ In reality, the time cost involved in later stages is not as substantial; the greater expense is in communication.}''} %- P17
\end{quote}
%P17: 我觉得最大挑战是在就需求确认那一块，因为有时候客户他把他的需求弄过来，他更多在意是功能的需求，其实在合规这方面需求他们有时候不要求，但是必须得做，所以说在需求这一块其实会有很大的的沟通成本之后，其实后面的时间成本其实不太多，更多是沟通成本。你会要求客户他为什么给他一些建议需要做，再就是我们自己如果他没有要求做到一些需求，我们可能需要给他额外的你做一些补充完之后再跟他交流。其实跟客户是有交流了，其实在跟开发有交流，有些开发商不知道合规为什么会不合规之类的，你还得跟他有额外的沟通交流成本。其实在需求确定那一块，其实成本是最大的，其实后面的话其实都比较小，因为你一开始做好了合规的有这方面意识，还一开始就做好了，其实后面都是很多的事情。

% Code
% 隐私合规整改的成本很高，整改时需要公司多部门、多团队支持，需要多角度考虑实现隐私合规	P3
    % P3: 最大难点其实是这个问题不是一个手段，一个机制就能把这个问题解决的，可能问题涉及的面会很广，处置或者说质疑的手段会很多，要从不同的角度去考虑，拉上不同的团队资源，然后得到老板的支持。等于你看到的是最早是4大类10个方面，然后后面又做了一些调整，做了一些增加，但是他每一句话都是需要去做相应的解读的，然后都可以去拆解出不同的问题下来，不同的要求下来，内部去做拆解，然后我就跟我们没有跟监管支持，我们根据监管意识的一些通报问题，然后去做裁决，开始对外将近100项，就有100多个至少100个控制点，这些点是需要去看的。

\para{Lack of fairness in accountability}
App engineers are unfairly considered to hold the major responsibility or at fault for privacy violations.
Upon the identification of privacy violations in an app, the app can be removed from app stores or notified by government. This could lead to a significant cost for the app providers, affecting their revenue and reputations.
Interviewees noted that when this happens, app engineers are often expected to be responsible for the privacy violations, without a fair discussion on factors such as business model and app design, that result in the privacy violations.
They mentioned that app providers impose penalties on the app engineers who develop or test of the codes that triggers the violations, such as reducing their salary, bonus, and job safety.
Moreover, government agencies only provide a short timeframe for app providers to rectify their apps.
To achieve compliance and release their apps, app providers pressure app engineers to resolve these issues quickly, threatening them with salary penalties, which ends up with developers working overtime. For instance, P17 shared:

\begin{quote}
    \textit{``The non-compliance issues in the app, which were not identified until reviews [by app stores], is actually quite serious within the company and will definitely result in lower performance evaluations.''} %- P17
\end{quote}
\vspace{0.5em}
\noindent\fbox{%
    \parbox{0.95\columnwidth}{%
    \textit{\jingtao{During SPRCs, app engineers encounter various challenges, such as inconsistencies in privacy review reports and testing criteria, constant updates to review specifications, and a lack of institutional support from app providers.}}    
    }
}
\vspace{0.5em}

% \subsubsection{Dilemmas of Vagueness for Privacy Compliance}

\subsection{RQ3: Solutions to Address the Challenges}
\label{subsec:solutions}

To deal with the challenges in Section~\ref{subsec:challenges}, interviewees developed a variety of strategies. 

\para{Performing pre-submission privacy certification}
\ignore{strengths and weaknesses of self-developed, and commercial apps}
App providers face both financial and reputational costs due to privacy violations in their apps, such as apps being removed from app stores or being publicly notified upon the identification of privacy violations.
%
%\ignore{App providers face both financial and reputational costs for privacy violations in their apps.
%
%Specifically, app stores require a significant amount of time to review submitted apps. Upon the identification of privacy violations by the stores, app providers must mitigate the violations and then resubmit the apps for another app store review, thereby indefinitely postponing the app release schedule.
%
%Additionally, for apps already listed on app stores, privacy violations can be detected through reviews initiated by government agencies and then publicized, causing severe damage to the app provider's reputation.}
%
Rather than being passively reviewed by external agencies, it is better for app providers to proactively discover privacy violations themselves.

\ignore{Given the fact that asking app developers to prove privacy compliance is infeasible due to their limited expertise (Section~\ref{subsubsec:insufficient}), some a}
App providers, exemplified by companies such as P1, P3, P6, P8, and P18, have adopted a pre-submission privacy certification process to reduce the chance of potential privacy violations,
\xw{Most of these companies are relatively large, for example, P1, P3, P8, and P18 all have 1,000+ employees.}
\ignore{which is frequent in large companies (like P1, P3, P8 and P18) who have large number of staff. \ignore{are more likely to frequently perform a pre-submission privacy certification.} }
This certification process, distinct from app developers' self-testing, often involves certifying the apps using professional and commercial certification tools from third parties, or conducting analyses within a dedicated security/privacy team.
The app provider will send their pre-release apps to a dedicated security team that monitors all privacy-invading system interfaces and reports their findings for developers to address.
P1 referred to this process as a ``privacy fallback'' or ``last-minute privacy self-review'' that catches as many issues as possible before the apps are publicized and become out of their control.
For example, P1\ignore{ and P6} noted that:

\begin{quote}
    \textit{``The product [app] was designed and assessed by the legal team. However, the codes might not always align with the design. Therefore, after compiling the final product, a security team is needed for a comprehensive test. %... the company does have clear standards, but adherence to these standards is not strictly enforced during the development or testing phases. Thus, a security department is needed to provide a safety net.
    ''}%- P1
\end{quote}

% P1：因为可能之前的阶段它都是一个偏君子协定，比如说法务评估了，产品设计了，他设计是按那样设计的，但的代码最后实现的不一定按他设计的来。所以等编译成最终的产物之后，需要有一个安全的团队做一个全盘的测试。:
% 对于前面在安全之前的这一些流程来说，是公司明确会有，但是它是一个规范，规范他遵不遵守，在他自己开发或者是测试过程中没有强制的去要求，所以需要一个安全部门去兜底，来在对他们的一些这些开发过程中。
% 以金钱换时间，购买第三方服务或者额外的工具开发人员来自测
% 兜底部门：对于前面在安全之前的这一些流程来说，是公司明确会有，但是它是一个规范，规范他遵不遵守，在他自己开发或者是测试过程中没有强制的去要求，所以需要一个安全部门去兜底，来在对他们的一些这些开发过程中。- P1

\ignore{\TODO{ignore?}
\begin{quote}
    \textit{``At ByteDance, for example, with every new version release, the relevant department is involved. They provide customized software to hook into your application and monitor which specific regulated interfaces are being called.''} - P6
\end{quote}
% 嗯，应该是会有专门的，据我了解，是会有的，会有定期的，奔驰我这不知道，但字节就是定期每处一个版本就会提交相关部门，部门就会提供定制的软件会去hook你的应用，调了哪些规定的接口 - P6
}

\para{Offering privacy compliance training sessions}
App engineers need more support from privacy and law professionals to better interpret the laws and eliminate confusion (Section~\ref{subsubsec:insufficient}).
Seven interviewees -- \xw{P5, P12, P13, P14, and P17 from relatively large companies with 1,000+ employees, and P6 and P7 from small companies} -- mentioned that their companies provided support through privacy compliance training sessions, led by either external or internal privacy law professionals.
\ignore{P5, P12, P13 and P14 work for relatively large companies with 1000+ employees, and P6 and P7 work for small-sized companies.}
\ignore{which indicates that the size of the companies has no influence on the privacy compliance training sessions offering. } 

P13 and P14 mentioned that their respective companies would invite officials from government agencies, including affiliated research institutions, to conduct training sessions for their employees.
Given that the rationale behind privacy laws and guidelines is best understood by the officials who draft them, the training provided by the officials offers the most authoritative information for privacy compliance.
This is confirmed by P13, who stated that these training sessions are effective since the officials would dissect the privacy compliance requirements, elaborate on the reasons for having them, and clarify who is responsible for meeting them, etc. 
Specifically, P13 mentioned that:

\begin{quote}
    \textit{``they [the officials] actually break down and analyze every aspect for you. They present all the details, essentially laying out the standards and how they operate.''}% - P13
\end{quote}
% 【Quote？】gentleking13: P13: 因为在这一点的话，实际上管局在宣传的时候，实际上都给你分析的条条框框，把（问题）全部给你拆出来，全部喂给你听，相当于是他们的标准是什么，他们那会是怎么样的，而我们这样的话也是觉得很不错的，因为这里面所有的每一个款项都给你拆解出来，非常的让你能够理解这其中的标准为什么要这么做，由谁来做。

%\begin{quote}
   % \textit{``We typically invite experts from institutions such as the CAICT \ignore{[China Academy of Information and Communications Technology]} or the CTTL [China Telecommunication Technology Labs], [both affiliated with MIIT].''} - P14
%\end{quote}

Six interviewees (P5, P6, P7, P10, P14, and P17) mentioned that their respective companies provide internal privacy compliance training for all employees.
According to the interviewees, the training takes various forms and spans the entire duration of employees' service. Some companies incorporate past cases of privacy violations into the training, and integrate the training into the onboarding process of new employees, to ensure that they are privacy-prepared before commencing their actual duties.
Additionally, these companies organize regular privacy trainings for product and engineering teams in order to help employees stay informed about new privacy updates.
These internal training sessions are customized to meet the specific business needs of each company. Therefore, interviewees often feel that they get better awareness of privacy compliance after the training, like what P14 said:

\begin{quote}
    \textit{``We also provide training during the onboarding of new employees... we conduct regular security training sessions related to compliance... We ensure timely synchronization if there are new standard requirements.''} %- P14
\end{quote}

\para{Collective sense-making on privacy compliance}
Since all the apps in the app stores have to be reviewed for privacy compliance, developers for different apps have to go through the same privacy review process, fostering collective sense-making and knowledge sharing among privacy engineers. Additionally, achieving privacy compliance involves the interplay of multiple domains such as engineering, privacy, and legal. It has become a challenging task for which no individual possesses all the knowledge required for resolution. Consequently, a community with different stakeholders in privacy compliance has been formed, including app developers, law experts, developers working for third-party certifiers and developers working for app stores. Such communities enable app developers to collaborate with each other and with other types of stakeholders in making sense of privacy violations. This process is supported by online social media groups (e.g., on WeChat) and forums.

Five interviewees (P1, P8, P11, P12 and P13) have actively participated in collective sense-making through various means. 
\xw{Interestingly, all five of these interviewees work for relatively large companies with 1,000+ employees, which potentially suggest that employees from these companies are more active in sharing knowledge.}
\ignore{All these five interviewees work for relatively large companies with 1,000+ employees, which indicates that the large-sized company frequently shares valuable information.} Firstly, after addressing a privacy violation, they often publicly share information about their privacy violation and the techniques used to address it. This aids others facing similar privacy issues in quickly identifying potential solutions.

Secondly, when confronted with a newly reported violation lacking a straightforward solution, they would post the violation on social media groups or forums. Members on these platforms then engage in discussions, offering potential solutions in response to the post. In most cases, these discussions lead to effective resolutions swiftly.
In addition to asking questions, they also participate in discussions about the violations of other companies.

Thirdly, they share updates on privacy review specifications and discuss the technical implications of these updates. These communities have become an essential, if not the only, source for small-sized companies to receive practical guidance for privacy compliance. For example, P12 told us:

\begin{quote}
    \textit{``I think this group is very useful. Whenever anyone has questions, I've noticed that everyone is quite enthusiastic. \ignore{The reason why some people might choose to copy others—by simply asking how others do it and then copying directly—is because this group makes it convenient to solve problems.} You can just raise your question, and everyone can offer solutions based on your situation. If you were to ask other companies [instead of asking in the groups], there might be delays depending on their availability, the companies may hold back information, or they ask for a fee.''}
\end{quote}

\para{Building privacy compliance into SDLC}
Interviewees highlighted that building privacy compliance into the software development life cycle (SDLC) helps avoid privacy violations. This will avoid app developers being soly responsible for privacy noncompliance results. 
Rather than relying primarily on app developers' implementation\ignore{ (as reported by P8 and P9)}, six interviewees (P1, P7, P11, P12, P13, and P18) reported that their companies involve two or more departments in different phases of SDLC in order to achieve privacy compliance.
\xw{We did not observe any notable differences caused by company size, since the interviewees are from both small and relatively large companies.}
\ignore{P1, P11, P12, P13 and P18 work for large-sized companies (1000+ members) while P7 works for small-sized company (50+ members), which indicates that both samll and large-sized companies attach importance to privacy compliance in SDLC.}
They emphasized the importance of testing privacy during the software testing phase, and noted that they consider privacy in the app's feature design phase, which allowed their company to avoid privacy violations at the early stage of software development, rather than having to reactively respond to reported violations. For example, P18 said:
%
%在开发app过程中，更好的建议是涉及app开发的所有部门都需要考虑自己的工作是否符合隐私合规。P1，P7，P11，P12，P13，P18都表示，在自己公司开发app过程中，有两个及以上的部门思考过如何实现隐私合规。P13表示在app功能设计时，需要考虑app收集用户信息的必要性，P4认为在app开发时，需要考虑获取用户信息的频率，P12则强调需要在测试阶段针对隐私合规进行测试，法务部门对隐私政策的评估。全流程实现app的隐私合规带来的好处，不仅是最终实现的app有尽可能少的隐私合规问题，还能提高全公司对谨慎收集用户敏感数据的意识，避免带来后续的政府通报。这种解决方法不可避免的会带来额外的工作，但相对于事后被检测出再整改来说，这种增加量是可以接受的。

% P13: （肯定态度）受访者认为提前做足隐私合规可以避免问题发生。	P13

\begin{quote}
    \textit{``[Privacy compliance] is considered in all steps. We focus on ensuring that any changes in documentation are coordinated with development. %\ignore{For instance, if development needs to integrate a new third-party SDK, the product team also ensures that the design of the related documentation complies with regulations, and checks whether the documentation needs to be updated. It also verifies whether the business requirements are consistent with the permissions collected and whether they are the minimum necessary.}
    ... Testing, whether using previously employed methods or new ones, is conducted to ensure consistency with policies and regulations.''} %- P18
\end{quote}

\vspace{0.5em}
\noindent\fbox{%
    \parbox{0.95\columnwidth}{%
    \textit{\jingtao{To address the challenges in SPRCs, app engineers perform pre-submission privacy certification, participate in privacy compliance training sessions, foster collective sense-making and knowledge sharing, and integrate privacy compliance into the SDLC.}}    
    }
}
\vspace{0.5em}

\subsection{RQ4: Positive Changes and Concerns}

Here we summarize several positive impacts that interviewees are generally agreed upon, and the remaining public concerns with the privacy law enforcement campaigns. 

\subsubsection{Positive Changes}
\label{subsubsec:positive}

\para{Reduction of privacy-invading apps}
Despite the challenges discussed in Section~\ref{subsec:challenges}, \xw{six interviewees (P1, P2, P3, P7, P11, and P12) personally felt that} the enforcement of privacy laws has reduced the number of privacy-invading apps and restricted the abuse of sensitive user data. 
\ignore{a notable number of interviewees (P1, P2, P3, P7, P11, and P12) clearly stated that, overall, the enforcement of privacy laws has effectively reduced the number of privacy-invading apps and restricted the abuse of sensitive user data.}
\xw{P11 mentioned that, prior to SPRCs, app providers, regardless of the types of their apps, attempted to collect as much user data as possible to build complete user profiles, and such data collection is now ``not as common as before''.}
\ignore{P11 emphasized the remarkable success of the enforcement efforts. Prior to enforcement, privacy-invading apps were common, and app providers could generate a complete user profile based on personal data collected. However, the current enforcement regulates apps' use of privacy APIs for collecting sensitive data from users and thus makes data misuse or abuse difficult.} 
P3 shared his before-SPRCs experience participating in the development of a flashlight app that aggressively accessed users' calendars and subscribed them to unauthorized charges. He highlighted that today's users can \xw{feel more assured}\ignore{rest assured} due to the enforcement of privacy laws.
Furthermore, from the perspective of an average user, P12 detailed three concrete positive changes resulting from the enforcement of privacy laws: 1) app privacy policies are becoming clearer and more comprehensive, 2) app user interfaces provide improved control over privacy (e.g., through runtime data collection requests and toggles for personalized ads), and 3) banking apps utilize secure keyboards to safeguard user input. 
\ignore{First, apps' privacy policies are now clearer and more comprehensive than ever before, providing users with transparency regarding the personal data collected and informing them of the means to exercise their individual rights. Second, app user interfaces are updated to allow users control over privacy-sensitive functionalities, such as sending runtime request prompts for data collection and disabling personalized ads. With these user interface updates, users feel more at ease about their privacy. Third, all banking apps now employ secure keyboards to protect user input, including credit card numbers and passwords.
For example, P12 mentions that:}
Specifically, P12 mentioned that:

\begin{quote}
    \textit{``First and foremost, you can see that privacy policies have been clearly laid out, which includes complaint channels, feedback methods, processing times, and collected information, allowing users to understand the policies at a glance... Secondly, from a user experience perspective, the requests for permissions and the ability to toggle features such as personalized ads... Then, for financial apps, the secure keyboards are much safer to use than the standard system keyboards, right?\ignore{ Regarding passwords, whether could be changed, whether the complexity required or whether simple or complex, there are specific requirements (from privacy laws).}...Thus, with the introduction of national laws, there's no doubt that things will continue to improve.''} %- P12
\end{quote}

\xw{The participants' positive feelings about the reduction of privacy-invading apps resonate with recent app privacy reports from independent agencies~\cite{2022reportprivacy,2021reportprivacy} and academic research~\cite{kollnig2023privacy}, which, \xwnew{for example, note a decrease in apps collecting device identifiers such as IMEI and MAC, and an increase in apps obtaining user consent for data processing since the launch of SPRCs.}}

\para{Growing agreement on the significance of privacy among app engineers}
By informing app engineers about privacy-violating apps and compelling them to address these issues, the enforcement of privacy laws exposes app developers to essential privacy principles. This helped to cultivate a consensus on the importance of safeguarding user privacy.
Not too long ago, stakeholders in mobile apps perceived the collection of user data as commonplace, exemplified by the statement made by the CEO of Baidu in 2018, suggesting that Chinese users were willing to trade privacy for convenience, safety, and efficiency~\cite{publicprivacy}.
Seven interviewees (P1, P3, P5, P11, P12, P14, and P16), however, emphasized their commitment to the ``minimum necessary rule'', a fundamental enforcement requirement, to ensure that they minimize data collection in alignment with their specific business needs whenever it occurs.
P7 observed a change in the mindset of app developers regarding the enforcement of privacy laws. Initially, some developers resisted addressing privacy compliance issues because of the additional engineering work involved. However, they later acknowledged the significance of privacy compliance, citing ``\textit{what industry leaders should do}'', which resulted in quicker responses to such issues:

\begin{quote}
    \textit{``Resistance to privacy compliance and rectification actions did exist, but the situation has improved now... They [app developers] resisted because too much work to do... 
    Most developers focus on their own staff, but privacy compliance is industry leader's considerations and is critical...
    \ignore{(Q: Why? ) Because it's become a necessity. (if not addressed,) you're already on the national inspection. Most developers primarily consider their own (perspective), but there's always a higher leader to consider, which inevitably encompasses broader aspects (to realize privacy compliance). }
    Nowadays, developers are capable of understanding privacy compliance. They can quickly provide feedback on the issues, which are then promptly addressed, and they are even willing to engage in communication.''} %- P7
\end{quote}

\para{Ongoing adaptation of privacy enforcement increases in-depth compliance}
Privacy law enforcement is continually being strengthened to promote comprehensive privacy compliance, with the incorporation of more stringent rules that extend into previously uncovered user cases. 
According to P4 and P18, new enforcement rules are introduced to achieve both privacy and usability simultaneously. 
Apps are now mandated to request explicit consent for using privacy-sensitive permissions. These requests must include an option to cancel, provide clear explanations for the necessity of the permission through non-deceptive messages, and be presented in a manner that does not disrupt the user experience.
P16 highlighted that privacy compliance is expanding its scope to cover business flows between apps and third parties. 
In cases where apps redirect users to external web pages and user data is collected on these pages, new rules require apps to inform users of the data collection before users can navigate to such web pages: 

\begin{quote}
    \textit{``If one of our services incorporates a function from a third-party contractor, and that contractor's interface requires users' personal information without a clear notification popup, \ignore{this could lead to a report. Users need to be informed. For example, in a banking app, there might be a button that, when clicked, opens a webpage asking the user to fill in some information. If this webpage is third-party and we don't provide any prompts to the user about this, }it would be considered a violation...\ignore{ Similarly, if users enter their phone numbers, and these numbers might be shared out,} it is mandatory to clearly inform them about such actions.''} %- P16
\end{quote}
% P16：之前可能年前一段，比如说前两年之前检测过，隐私政策之前获取一些权限，然后最近这两年或者说最近的话它深入到业务了，比如说我们的某个业务，然后接入了第三方的一些东西，接入了第三方厂商，或者说第接入了第三方的外包商，然后在外包商的界面提供了需要提供让用户提供一些个人信息，然后没有一些明显告知的一些东西，他会一些通报。更多的是告知用户，就是说比如说我在我的银行界面，然后在我的银行一个有个按钮，然后点按钮之后，用户在VIP打开了一个VIP界面，就是打开了一个网页，打开了一个网页，需要用户填一些信息，其实这个网页是第三方的，需要用户填一些信息，然后如果在这个页面里面，我们没有对用户一些提示的话，就算是有违规的行为的。比如说你输入手机号码，你可能这个手机号码会出出去，必须要很明显的告诉他（收集手机号的行为）。

Moreover, third-party SDKs were initially viewed as black boxes, posing challenges for app developers in addressing privacy violations within them. 
The evolution of privacy enforcement has streamlined the handling of these SDKs. In particular, government agencies began directly detecting privacy violations within the SDKs and reporting them to the SDK providers. 
As indicated by the interviewees (P4, P7), this change was effective: many third-party SDKs are now proactively pursuing compliance similar to apps, and developers feel less concerned about using these SDKs.

% \begin{quote}
%     \textit{``''} - PX
% \end{quote}

% 1）Positive(有积极意义的):
    % 政府政策对隐私合规的标准正在逐渐变得严格	P4
    	% P4：对的前两年都没有这么严的，你知道吧，就最近三年时间。对，最近三年开始慢慢严，然后就越来越详细，搞的规则也特别多，比方弹窗都是近几年开始加的，原来都没这个时候没这个说法。
    % 隐私合规的检测之前仅仅对功能性是否合规进行检查，现在逐渐便严格，会对检测具体业务是否合规。打开的页面，需要明确告知用户页面的功能	P16
    	% P16：之前可能年前一段，比如说前两年之前检测过，隐私政策之前获取一些权限，然后最近这两年或者说最近的话它深入到业务了，比如说我们的某个业务，然后接入了第三方的一些东西，接入了第三方厂商，或者说第接入了第三方的外包商，然后在外包商的界面提供了需要提供让用户提供一些个人信息，然后没有一些明显告知的一些东西，他会一些通报。更多的是告知用户，就是说比如说我在我的银行界面，然后在我的银行一个有个按钮，然后点按钮之后，用户在VIP打开了一个VIP界面，就是打开了一个网页，打开了一个网页，需要用户填一些信息，其实这个网页是第三方的，需要用户填一些信息，然后如果在这个页面里面，我们没有对用户一些提示的话，就算是有违规的行为的。比如说你输入手机号码，你可能这个手机号码会出出去，必须要很明显的告诉他（收集手机号的行为）。
    % 隐私合规相关的检测标准也在一直变化		P18
    	% P18：当然我们定的一些都是按照国发国家规定，工信部来的是不管第三方机构还是音乐上架一些市场都是以那种作为依据，而且也不是不变的，可能你很早的之前权限收集，他没有要求，后面他可能有要求变化的，变化，有要求的话我们就把它加进去。对新的而且APP有些东西也是一直升级。现在要反正时间的话应该有个跨度，频繁大概不频繁。但是它的趋势是反正越来越严格，越来越注重用户的隐私，之前很多一些权限越来越收紧，反正系统也是的，之前可能是很开放的，后面可能有什么权限弹框告知用户之类的，反正权限会越来越越来越严格。
% 对于第三方SDK开发商的限制
    % 大型第三方SDK开发商都会积极按照政府要求来开发实现	P4
    	% P4：他们这种你接的第三方SDK都是一些大厂，他们不会一般就是说上面的政策就是国家的政策不改的话，他们一般是不会改不会变动一些，不会去变做变动，会去长时间做变动，就好比说这一款SDK如果发布了之后，用户去使用它，没有如果隐私会隐私合规的一些风险的话，他们是后面是不会变动的。
    % 大公司开发的SDK很少有合规问题	P7
    	% P7：大厂商的基本上都是没问题的。一些可能去对对接的小厂商，实在不行就不用了是吧

\subsubsection{Remaining Concerns}

\para{\xwnew{Potential techniques to circumvent SPRCs}}
The major concerns, as expressed by six interviewees (P4, P9, P10, P12, P15, P17), are the potential for privacy-invading apps to circumvent privacy enforcement through various means.
P9 and P15 noted that, owing to the open nature of the Android platform, privacy-invading apps can be distributed with any ways (such as a downloadable link posted online) other than app stores that require privacy compliance. These apps can still impact a significant number of victim users, e.g., through referrals of popular social media platforms. 
P17 referred to the scandal of the Pinduoduo app, the most used C2B e-commercial app in China, that collects user data by exploiting operating system vulnerabilities~\cite{pinduoduo}, and speculated that similar bypass techniques might be employed by other companies without being detected. 
P10 reported that it is a common practice for large companies to support cloud-side configuration in their apps. During app reviews, they disable privacy-invading behaviors using this configuration, and only re-enable them after receiving approval from app stores (similar to~\cite{lee2019understanding}). Specifically, P10 mentioned that:

\begin{quote}
    \textit{``This approach [bypassing detection] is widespread. For example, most companies employ a strategy where they place settings in the cloud... They deactivate these settings when the apps are under review, and then once the app passes the store's review, they reactivate them.\ignore{ The toggle involves reading a configuration from the server. Initially, it is set to false, ... and it will request permissions after the user has agreed. Once the app goes live, they change it to true, applying for these permissions before you have agreed.}''} %- P10
\end{quote}

% 2）Negative(还需要完善的):
% 隐私合规检测可能会有遗漏
    % 应用商店人工渠道审核并不严格，可能有漏过的检测的点		P7
    	%（有哪些可能漏过的检测？）刚刚说的隐私政策里面有一个推送。因为推送这个点需要人点进去看。因为人点进去看，有些应用需要登录之后相应的功能点才会开放。然后加上分发渠道，就各个发布渠道，发布的APP比较多，有的时候它其实人工审核它跳过了，能够感受到它跳过了。我估计是抽检的行为。
%P7：因为是人工审核，也就因为在各个分发渠道，其实人工审核不严的，因为人工审核太多了，他毕竟在分发渠道，他每天（上架应用太多）。

% 隐私合规检测可能会被绕过
    % APP不在应用商店上架导致检测绕过	 P15
    % 应用开发商存在对隐私合规检测绕过的行为，例如拼多多；还有国家可能也检测不出来的一些行为	P17
    	%P17：当然整体的隐私合规的框架肯定是以国家标准为准，其他国家那都覆盖到了。明白，只是有些点因为国家也不能检测出，有时候它一些最新的检测不出来。当然这个肯定类似于拼多多，然后其实很多的但是其他的平台可能没爆雷，只是倒是有一些小道消息都是说也有，当然这个就也不能多说对吧？其实整个其他的如果在明面上的，其实很多都是通过那种，他其实一开始就有那些企业之类全都写好了的，当然也类似于那种之前有举例，就是说 QQ，然后QQ它有一些协议里面就说了，比如说QQ它的所有权归腾讯所有，这些其实都是他有自己的一些协议里面已经写好了的，主要是看用户有没有仔细去看，但是他肯定协议里面都写好了的，当然大家尊重合同精神对吧？然后人家已经写好了，但已经同意了，然后肯定以协议为主，明面上肯定大家都这样做，但私底下的话可能大家有一些其他自己的做法。
    % APP不上架应用商店，会通过聊天群，分享平台等，通过链接的方式下载，对于这种APP，并不能有效进行隐私合规监管，这种APP也不会在意隐私合规		P9
    	%P9：包括现在，在这种整改的一些情况下，目前我发现的一些特点就是有很多c端软件或者说是对有很多c端软件基本就是通报就通报，然后我也不在乎，应用市场也不上架，我的推广靠微信流量，靠一些网站就ok了。我直接扔给你链接，然后你微信一点击，打开浏览器下载，微信不支持下载，我就浏览器打开，浏览器玩深一点就下载下来就安装上了。（就不能）检测。这个东西你要是能检测那些黄片APP咋上架，那咋流传呢？ 对现在不法分子，基本有说他正常正经在应用市场上架，不都是通过浏览器引流，腾讯先从微信上，然后之后整一批意向用户在引流在推广，发点朋友圈是吧？那就可以了。 其实他就规范不了，他真正想规范的，规范的基本都是一都是想规范的，不想规范它也规范不了。
    %  应用开发商可能使用绕过技术，通过将不合规的功能放在云上，在检测的时候关闭相关功能，在非检测的时候再打开	P10
    	%P10：这个（使用技术绕过检测）也很多。就很简单，举个例子，大部分公司都这样做的，他说搞一些开关配置放在云上面，然后你做合规的那一天他把这个东西关掉，然后等商店审核过了，他就把它打开。（开关是指）就是去读一个配置，你去你服务器上读一个配置，这样子就可以了。但是他之前是false，然后他false的时候他就会在你用户同意了之后，他再去申请，然后等你上线的时候他就把它改成true，默认无论你的这些权限或者去申请这些权限，你同意之前。
    % APP会通过获取缓存中的内容来规避获取频次的问题	P5

\para{\xwnew{Manipulation of privacy policies may lead to false compliance}}
\ignore{Another way to evade privacy enforcement is through manipulating privacy policies with deceptive statements.}
When an app collects unnecessary sensitive user data, privacy regulations prohibit it to ensure compliance.
However, P4, P12, and P15 highlighted instances where app providers continued collecting data. They added seemingly reasonable but deceptive statements to the app's privacy policies to justify the need for such data collection.
In most cases, the app can still be published on app stores since the stores essentially permit any privacy practices listed in the privacy policies, as long as users agree.
This manipulation of privacy policies is due to either innocent app developers who copy the privacy policies of other compliant apps, hoping it helps them become compliant as well, or malicious parties unethically manipulating privacy policies. For instance, P4 said:

\begin{quote}
    \textit{``Platforms [app stores] will not detect this [excessive data collection described in the privacy policy]. Because from the platform's perspective, \ignore{they will only assume that you have already marked it in the privacy policy, and when a user has the app and clicks to confirm, }they assume that the user has read that privacy policy. Yes, that's the standard procedure. As long as they [app stores] know there won't be any legal risks for the platforms, they generally allow it.''} %- P4
\end{quote}

\vspace{0.5em}
\noindent\fbox{%
    \parbox{0.95\columnwidth}{%
    \textit{\jingtao{\xwnew{Interviewees report that} SPRCs bring several positive changes, such as the reduction of privacy-invading apps, growing agreement on the significance of privacy among app engineers, and increased compliance due to the ongoing adaptation of privacy enforcement. However, app engineers are also concerned about the presence and use of techniques to evade privacy enforcement.}}    
    }
}
\vspace{0.5em}

\section{Discussion}
\label{sec:suggestions}

\ignore{In this section, we present several general suggestions based on the challenges identified in the SPRC-based privacy enforcement campaigns. We also discuss the implications of this research to the enforcement of privacy laws of other countries.}

\ignore{
We hope these suggestions can benefit future campaigns, and inform the other approaches to enforcing privacy laws. 
}

\para{Suggestions for SPRC-based privacy enforcement}
\xw{SPRCs are large-scale campaigns under the directives of central authorities that involve various stakeholders, such as app providers and app stores, in the privacy enforcement process.}
\xw{As illustrated in Section~\ref{subsubsec:positive}, several app engineers reported positive changes based on their experience and perceptions, such as in reducing the explosion of privacy-invading apps and building agreement on the significance of privacy among app engineers.}
However, these positive changes do not come cheap.
In this study, interviewees frequently reported that enforcement pressure has unfortunately been largely shifted onto app developers. These developers often find themselves ill-prepared for the diverse range of tasks they are now required to handle, such as studying privacy laws, communicating, and conducting compliance self-testing. The root cause of these difficulties lies in the absence of a robust mechanism (or process) within app providers to effectively coordinate (human) resources according to their expertise in privacy compliance.
For example, offloading the task of studying laws to legal teams and the responsibility of communication to program managers, thus relieving app developers of these efforts.
In the long term, it would be beneficial for app providers to establish a formal and sustainable process that guides multiple departments to collaborate seamlessly in order to ensure privacy compliance. 

Furthermore, government agencies have successfully motivated app providers to address privacy violations through reporting and imposing heavy penalties, such as app removal from stores. However, resolving these violations requires a thorough understanding of privacy laws. While government agencies are the primary source of information for app providers, this study suggests they should have gone extra mile educating especially smaller companies. While larger firms may host training sessions with government officials, smaller companies, like P9's, often lack resources for such initiatives. Therefore, increasing accessibility to these sessions would greatly benefit smaller companies.
Also, while SPRCs benefit from scalability and rapid sanctioning, they lack the procedural safeguards of privacy laws in other countries (e.g., those involving court procedures). Adding mechanisms to enhance assurance for companies facing incorrect or unfair judgments during sanctioning could be beneficial.

\ignore{Furthermore, the government agencies have achieved great success in motivating app providers to address their privacy violations by reporting them and imposing heavy penalties (e.g., terminate their services by removing apps from app stores). 
However, resolving these violations requires a comprehensive understanding of privacy laws and regulations. 
The government agencies are considered the most authoritative source of information on this matter for app providers. 
However, this study suggests that the agencies should have gone the extra mile in educating app providers, especially small- or medium-sized companies. As mentioned in Section~\ref{subsec:solutions}, larger companies may invite government officials to conduct training sessions. However, smaller companies like P9's, which have limited resources, do not typically adopt such training sessions. Therefore, making these sessions widely accessible would greatly benefit these smaller companies.}

\para{Implications to the enforcement of other privacy laws}
SPRC-based enforcement represents an effort to enforce Chinese privacy laws with increased investment in administrative and public resources.
Applying SPRCs directly to the privacy laws of other countries can be challenging due to factors such as differences in governmental structure and available resources.
However, since the privacy laws of China and other countries share strong similarities~\cite{emmanuel2020china}, we believe that adopting at least some of the best practices from SPRC-based enforcement may help address certain persistent concerns in enforcing other laws.
An example of this practice is highlighting the responsibility of app stores to conduct app privacy reviews.
In some countries, app stores are not expected to ensure privacy compliance of apps through proactive privacy testing, but rather rely on self-compliance, for instance, apps self-claim their data practices via the Google Play's Data Safety section~\cite{googledatasafty}.
% "Provide information for Google Play's Data safety section": https://support.google.com/googleplay/android-developer/answer/10787469?hl=ens
%
Given that self-regulation has almost stagnated~\cite{amos2021privacy}, requiring app stores to launch systematic privacy reviews on their apps can be helpful.

\para{Limitation discussion}
\xwnew{This study focuses on app engineers and does not include other stakeholders involved in app privacy compliance, such as app stores, third-party privacy certifiers, and government agencies (as discussed in Section~\ref{subsubsec:stakeholders}). Therefore, our findings reflect only the perceptions of app engineers and may not fully capture the broader landscape.
For instance, insights into how app stores review popular versus unpopular apps, as well as their outsourcing of privacy reviews (as noted in Section~\ref{subsec:challenges}), are based on app engineers' personal perceptions and limited exposure to these processes. They may not have complete knowledge of app store review procedures or whether such popularity-based review criteria actually exist. Consequently, the accuracy of these insights should be further validated by involving app store representatives.
Nevertheless, our findings remain valuable. By focusing on the perspectives of app engineers, who are key stakeholders in app privacy compliance, including designing, testing, and implementing privacy controls, we establish an initial understanding of the problem space, and explore app engineers' understanding of SPRCs, the challenges they face, and their assessment of these concerns. Even if some bias exists, the fact remains that app engineers are impacted by SPRCs and face significant challenges in their roles, indicating the need for additional support to facilitate their work.
As a next step, we plan to conduct a multi-stakeholder exploration (e.g., through focus groups and co-design sessions) to generate a more comprehensive and balanced understanding of SPRCs.}

\xwnew{Additionally, the app engineers we interviewed may have self-censored their responses due to company policies, government oversight, or concerns about potential negative impacts on their careers, which could introduce bias into our findings. While we discussed various challenges and complaints reported by app engineers in Section~\ref{subsec:challenges} (e.g., inconsistencies in privacy review reports, frequent changes to app review specifications, and lack of institutional support), our findings may lack a more comprehensive and candid discussion of criticisms regarding SPRCs. For example, these challenges and complaints are primarily directed toward app stores, third-party privacy certifiers, and their own companies, with no direct criticism of the central government, despite the fact that SPRCs are government-initiated.
This lack of criticism toward the central government may be due to self-censorship, with app engineers potentially avoiding any negative commentary about the government out of concern for the potential impact on their careers. As a result, the self-reported data from these engineers may be incomplete and fail to fully reflect their true opinions about the government-led SPRCs.
During our interviews, we took several steps to mitigate this bias, such as anonymizing responses, ensuring confidentiality, building rapport and trust with interviewees, and asking behavior- or situation-based questions to encourage more open responses. 
Despite these efforts, the findings may still provide only a partial view of app engineers' perceptions of SPRCs.
Nevertheless, our study offers an initial glimpse into app engineers' views on SPRCs. We recommend that future research incorporate non-self-report methods, such as discourse analysis of public forums like Stack Overflow, to complement the interview findings and provide a more comprehensive perspective.
}

Our study is limited in scalability. Given the exploratory nature of the study, we adopted interviews with a small sample (18 participants) to probe open-ended and in-depth nuances, details and reasons behind app engineer's' opinions and practices in app privacy compliance. While this small sample is common in interview studies and reached thematic saturation, a larger sample is inevitably helpful to improve the scalability and representativeness of the findings. We acknowledge this limitation and hope to expand the scalability through methods like surveys in future.

\section{Related Work}
\label{sec:related}

\para{Research on the enforcement of privacy laws}
With privacy law enforcement comes the question of its outcomes. Previous research uses retrospective and comparative methods to address this.
\ignore{
With the enforcement of privacy laws comes the question of what outcomes or impacts the enforcement has led to.
Prior research addresses this question using both retrospective and comparative methods.
}
For example, one line of research, represented by GDPRxiv~\cite{sun2023gdprxiv}, Saemann et al.\cite{saemann2022investigating}, and Wolff et al.\cite{wolff2021early}, collects information on past privacy law violations, and then conducts aggregated analyses or case studies to gain a better understanding of these violations (e.g., what privacy principles were commonly violated).
Another line of research compares the apps, websites, or their privacy policies posted before and after the enactment of privacy laws~\cite{kollnig2023privacy,urban2020measuring,degeling2019we,linden2020privacy,amos2021privacy}, in order to measure the impacts of the enforcement of privacy laws.
These studies yield a series of observations.
For example, many more apps started to implement consent for data collection~\cite{kollnig2023privacy} and reduce the amount of data sharing~\cite{degeling2019we}.
Some studies show that privacy policies provide better transparency by covering more data practices with improved visual representations~\cite{linden2020privacy}, while others have slightly contending observations, i.e., privacy policies have doubled in size and become more difficult to read~\cite{amos2021privacy}.
Unlike the above studies that conduct retrospective or comparative analysis on the violation reports and apps, this study investigates the workflow, challenges, solutions, and overall results of the Chinese privacy law enforcement through the lens of app-related engineers, thanks to the deep involvement of them in the privacy compliance.

\para{Research on techniques to aid privacy compliance}
Both preventive and detective approaches have been developed in academia and industry to help achieve privacy compliance.
Examples of preventive approaches include visual and structured representations and modeling of privacy laws to guide organizations in privacy management~\cite{torre2019using, tom2018conceptual}, automated tools to perform GDPR-compliant operations on legacy systems~\cite{agarwal2021retrofitting}, an information flow tracking framework that supports privacy enforcement policies~\cite{klein2023general}, and tools for automatically generating compliant privacy policies based on the analysis of app behaviors~\cite{yu2015autoppg, zimmeck2021privacyflash}, etc.
On the other hand, detective approaches are mostly driven by reported noncompliant privacy practices. For example, many prior studies proposed techniques to identify violations caused by privacy policies through flow-to-policy analysis~\cite{zhao2023demystifying, andow2020actions, bui2021consistency, samarin2023lessons}.
Other techniques aim to address specific types of noncompliance, such as checking the absence of explicit and freely given consent before data collection~\cite{kollnig2021fait, koch2023ok, nguyen2021share, nguyen2022freely}.
In addition to that, commercial tools, such as Google Checks~\cite{checks}, Data Theorem Mobile Secure~\cite{data-theorem-mobile-secure}, AppCensus~\cite{appcensus}, and NowSecure Platform~\cite{nowsecure}, have been recently introduced with capabilities to check the accuracy of privacy labels that developers self-report on app stores.
Notably, none of the techniques or tools proposed so far claim to support full compliance detection for any privacy laws.
The participants in this study may not directly use these techniques or tools, but they reported using open-source or commercial tools designed with similar methodologies. 

\para{Research on the challenges to privacy compliance}
Previous research studies have extensively explored the challenges of achieving genuine privacy compliance in the process of EU/US privacy law enforcement, with most of them conducted through surveys or interviews with privacy stakeholders~\cite{klymenko2022understanding, horstmann2024those, sirur2018we, von2023we, senarath2018developers, alhazmi2021m, tahaei2021privacy, li2022understanding, alomar2022developers, ayalon2017developers, van2021data, bednar2019engineering}.
Major challenges identified in these studies include: 1) the disconnection between technical implementations and general privacy principles~\cite{klymenko2022understanding, sirur2018we, von2023we, alhazmi2021m, tahaei2021privacy}, 2) failures in the interactions between engineers and legal experts~\cite{bednar2019engineering, horstmann2024those}, and 3) a lack of knowledge about third-party SDKs being used~\cite{li2022understanding, alomar2022developers}, and etc.
The enforcement of Chinese privacy laws is based on SPRCs, involving a more extensive range of stakeholders and more frequent/complex interactions between them (which do not appear in the enforcement of US/EU privacy laws).
This allows us to report a set of unique challenges that manifest in SPRC-based enforcement~\ref{subsec:challenges}.

\section{Conclusion}
\label{sec:conclusion}

\xwnew{Recent years have seen significant positive changes due to the enactment of privacy laws. However, regulators are often under-resourced and have limited bandwidth to investigate the large number of apps.}
In contrast, since 2019, China has implemented Special Privacy Rectification Campaigns (SPRCs) to address widespread privacy issues in its mobile app ecosystem.
\xwnew{The campaigns feature large-scale privacy reviews of apps and impose strict sanctions for identified violations.}
Despite some reports on SPRCs, the effectiveness and potential issues of these campaigns remain largely unclear.
This paper seeks to evaluate this \xw{new campaign-style privacy enforcement approach} and offer insights for future law enforcement practices. Through 18 semi-structured interviews with app-related engineers involved in SPRCs, we report new understanding about their process, challenges, solutions, and the impact of these large-scale campaigns. Our findings reveal the operational workflow of SPRCs and the challenges faced by app developers in compliance with these campaigns. Despite developers' adoption of technical and behavioral solutions to overcome these challenges, concerns persist regarding the effectiveness of SPRCs in addressing all privacy violations.

\section{Acknowledgments}
%-------------------------------------------------------------------------------
We would like to extend our heartfelt gratitude to the shepherd and the anonymous reviewers for their invaluable insights and constructive comments on our work. This work is funded by the National Key Research and Development Program (No.2022YFB4501300), National Natural Science Foundation of China under Grants 62202194. This work was supported by Ant Group through CCF-Ant Research Fund. 

\bibliographystyle{plain}
\small{\bibliography{main}}

% \if0
\vspace{1cm}

% Jingtao: 
% Appendix A: Recruitment Info
% Appendix B: Interview Transcript
% \clearpage
\newpage
\appendices
\section{Participants' information}
Table~\ref{table: Participants' info} lists the background information of the participants in our interview.

\begin{table}[hb!]
% \begin{threeparttable}[]
\centering
\caption{Participants' Background}
\scriptsize
\label{table: Participants' info}
    \resizebox{\linewidth}{!}{
    \centering
    \begin{tabular}{ccccccc} 
    \toprule
    No. & Comp. Size & City & Years & App Type & Department & Role \\
    \midrule
    \rowcolor[gray]{.8}
    1 & 6,000+ & Beijing & 3+ & Entertainment & Development & Developer\\
    2 & 2,000+ & Wuhan & 3+ & Finance & Development & Team leader\\
    \rowcolor[gray]{.8}
    3 & 200,000+ & Hangzhou & 5+ & Safeguard & Law & Policy interpreter\\
    4 & 40+ & Shanghai & 6+ & Tool & Development & Developer\\
    \rowcolor[gray]{.8}
    5 & 4,000+ & Beijing & 5+ & Security & Security & Regulator\\
    6 & 100+ & Shanghai & 1+ & Car system & Development & Developer\\
    \rowcolor[gray]{.8}
    7 & 50+ & Hefei & 1+ & Finance & Security & Security tester\\
    8 & 5,000+ & Beijing & 4+ & Estate & Development & Developer\\
    \rowcolor[gray]{.8}
    9 & 20+ & Shenzhen & 3+ & Education & Development & Developer\\
    10 & \dag & Singapore & 6+ & E-commerce & Development & Developer\\
    \rowcolor[gray]{.8}
    11 & 10,000+ & Hangzhou & 1+ & Social, Office, Education & Security & Security tester\\
    12 & 3,000+ & Beijing & 2+ & Finance & Development & Security tester\\
    \rowcolor[gray]{.8}
    13 & 1,000+ & Wuhan & 3+ & Tool & Technology & software manager\\
    14 & 1,000+ & Suzhou & 4+ & E-commerce & Security & Security tester\\
    \rowcolor[gray]{.8}
    15 & 100+ & Hangzhou & 3+ & Game & Security & Security engineer\\
    16 & 20,000+ & Zhengzhou & 4+ & Finance & Security & Security engineer\\
    \rowcolor[gray]{.8}
    17 & 2,000+ & Changsha & 1+ & Communication & Test & Test engineer\\
    18 & 1,000+ & Hefei & 5+ & News & Development & Technical leader\\
    \ignore{\rowcolor[gray]{.8}
    19& ccc& ccc& ccc& ccc & ccc & ccc\\
    20& ccc& ccc& ccc& ccc& ccc & ccc\\}
    \bottomrule
    \end{tabular}
    }
    \begin{tablenotes}[]
        \item[1]\dag~~P10 preferred not to disclose the company size.  
    \end{tablenotes}
% \end{threeparttable}
\end{table}
% \fi
%with Privacy

\end{document}